\documentclass[useAMS,usegraphicx,usenatbib]{mn2e}

\newcommand{\aap}{A\&A}
\newcommand{\apj}{ApJ}
\newcommand{\apjl}{ApJ}
\newcommand{\apjs}{ApJS}
\newcommand{\araa}{ARA\&A}
\newcommand{\mnras}{MNRAS}
\newcommand{\nat}{Nat}
\newcommand{\icarus}{Icarus}
\newcommand{\pasp}{PASP}
\newcommand{\physrep}{Phys. Rep.}
\newcommand{\pre}{Phys. Rev. E}

\title[Primordial protostellar systems]{Formation and evolution of primordial protostellar systems}

\author[Greif et al.]{\parbox{17.5cm}{Thomas H. Greif$^1$\thanks{E-mail: tgreif@mpa-garching.mpg.de}, Volker~Bromm$^2$, Paul~C.~Clark$^3$, Simon~C.~O.~Glover$^3$, \newline
Rowan~J.~Smith$^3$, Ralf~S.~Klessen$^3$, Naoki~Yoshida$^4$ and Volker~Springel$^{5, 6}$\vspace{0.3cm}}
\\$^1$ Max-Planck-Institut f\"{u}r Astrophysik, Karl-Schwarzschild-Stra\ss e 1, 85740 Garching bei M\"{u}nchen, Germany
\\$^2$ Department of Astronomy and Texas Cosmology Center, University of Texas, Austin, TX 78712, USA
\\$^3$ Zentrum f\"{u}r Astronomie der Universit\"{a}t Heidelberg, Institut f\"{u}r Theoretische Astrophysik, \\ Albert-Ueberle-Stra\ss e 2, 69120 Heidelberg, Germany
\\$^4$ Institute for the Physics and Mathematics of the Universe, University of Tokyo, 5--1--5 Kashiwanoha, \\ Kashiwa, Chiba 277--8568, Japan
\\$^5$ Heidelberg Institute for Theoretical Studies, Schloss-Wolfsbrunnenweg 35, 69118 Heidelberg, Germany
\\$^6$ Zentrum f\"{u}r Astronomie der Universit\"{a}t Heidelberg, Astronomisches Recheninstitut, \\ M\"{o}nchhofstr. 12-14, 69120 Heidelberg, Germany}

\begin{document}

\maketitle
\topmargin-1cm

\begin{abstract}
We investigate the formation of the first stars at the end of the cosmic dark ages with a suite of three-dimensional, moving mesh simulations that directly resolve the collapse of the gas beyond the formation of the first protostar at the centre of a dark matter minihalo. The simulations cover more than $25$ orders of magnitude in density and have a maximum spatial resolution of $0.05\,{\rm R}_\odot$, which extends well below the radius of individual protostars and captures their interaction with the surrounding gas. In analogy to previous studies that employed sink particles, we find that the Keplerian disc around the primary protostar fragments into a number of secondary protostars, which is facilitated by H$_2$ collisional dissociation cooling and collision-induced emission. The further evolution of the protostellar system is characterized by strong gravitational torques that transfer angular momentum between the secondary protostars formed in the disc and the surrounding gas. This leads to the migration of about half of the secondary protostars to the centre of the cloud in a free-fall time, where they merge with the primary protostar and enhance its growth to about five times the mass of the second most massive protostar. By the same token, a fraction of the protostars obtain angular momentum from other protostars via N-body interactions and migrate to higher orbits. On average, only every third protostar survives until the end of the simulation. However, the number of protostars present at any given time increases monotonically, suggesting that the system will continue to grow beyond the limited period of time simulated here.
\end{abstract}


\section{Introduction}

The formation of the first stars, also termed Population~III (or Pop~III), heralded an end to the physical simplicity of the Universe at the end of the cosmic dark ages \citep[e.g.][]{bl01,bl04a,cf05,glover05,bromm09,loeb10}. Their radiation heated and ionised the intergalactic medium \citep[IGM; e.g.][]{bkl01,schaerer02,abs06,awb07,jgb07,yoshida07,whalen08a,greif09b}, and the violent supernova explosions that marked their deaths enriched the primordial gas with the first heavy elements \citep[e.g.][]{heger03,un03,wa08b,greif10}. One of the main goals is therefore to derive their characteristics, such as their initial mass function (IMF) and rotation rate, which govern their influence on the IGM and subsequent stellar generations.

With the exception of magnetic fields, whose strength are poorly understood, the initial conditions for this problem are well-posed, since the fluctuation power spectrum and elemental composition of the gas after recombination have been constrained by observations of the cosmic microwave background \citep[CMB; e.g.][]{komatsu09}. However, the complex interplay of the relevant physical processes over many orders of magnitude in spatial and temporal scale have so far precluded a comprehensive answer. Nevertheless, pioneering one-dimensional calculations as well as three-dimensional numerical simulations have suggested that the first stars formed in dark matter (DM) minihaloes of mass $M_{\rm vir}\sim 10^5-10^6\,{\rm M}_\odot$ at redshifts $z\ga 20$ due to the formation of molecular hydrogen \citep{htl96,tegmark97,abn02, bcl02}. The virialised gas cools to a minimum temperature of about $200\,{\rm K}$, which is set by the internal rovibrational transitions of H$_2$ \citep{on98}. The gas cloud then becomes Jeans-unstable and undergoes runaway collapse, resulting in the formation of a protostar at the centre of the halo \citep{yoshida06b,yoh08}. In the absence of fragmentation, an extrapolation of the high accretion rate within the Kelvin-Helmholtz time yields a characteristic mass of $M_{\rm *}\ga 100\,{\rm M}_\odot$ once the star reaches the main sequence, two orders of magnitude higher than for present-day stars \citep{kroupa02,chabrier03}.

While providing a plausible estimate of the expected mass scale, this argument glosses over details of the collapse and accretion process. In particular, the Courant constraint limited the simulations to a narrow timespan centred on the formation of the first protostellar core. Fragmentation therefore only appeared if the collapse of a secondary clump was synchronised with the first \citep[e.g.][]{tao09}. Furthermore, the formation and possible fragmentation of a circumstellar disc could not be followed. Recent studies have circumvented this limitation by employing `sink particles' that represent growing protostars \citep[e.g.][]{bbp95,kmk04,jappsen05,federrath10}. Irrespective of the scale on which the sink particles were inserted, these studies found that the protostellar disc fragmented strongly into a group of protostars, due to the high accretion rate from the envelope on to the disc, and the very efficient cooling of the disc by H$_2$ lines and collision-induced emission (Clark et al. 2008, 2011a,b; Stacy et al. 2010, Greif et al. 2011a). This led to the inference that Pop~III stars may have formed with a range of masses, and that the final mass function depends on the efficiency of fragmentation and merging.

Unfortunately, the utilization of sink particles also comes at a cost. In particular, the boundary conditions imposed on the gas near the accretion radius are not a self-consistent extension of the underlying hydrodynamic equations. With the exception of supersonic inflows, where the trajectories become ballistic, the boundary between the sink particles and the surrounding gas may be subject to numerical artefacts. The complicated torques near the accretion radius are not captured accurately, which might artificially enhance or prevent fragmentation. The loss of resolution on the scale of the accretion radius also sets a minimum scale on which fragmentation may occur. Finally, close encounters between sink particles may result in dynamical ejections (e.g. Greif et al. 2011a; Smith et al. 2011a,b), although the orbital energy may in fact be dissipated through unresolved torques. An accurate account of fragmentation and protostellar interactions on scales comparable to the physical sizes of the protostars therefore requires a more self-consistent treatment.

In the present study, we address these issues by performing three-dimensional, moving-mesh simulations of the formation and evolution of primordial protostellar systems in minihaloes. Similar to \citet{greif11a}, we start from cosmological initial conditions and follow the collapse of the gas up to the formation of the first protostar. Once the collapse stalls, we deviate from previous work and do not insert sink particles. Instead, we self-consistently evolve the chemistry and hydrodynamics of the gas beyond the `Courant myopia' of the first protostar to capture the formation and fragmentation of the circumstellar disc, as well as the evolution of the nascent protostellar system for many dynamical times. The maximum spatial resolution of $0.05\,{\rm R}_\odot$ allows us to resolve the immediate vicinity of the protostars and their complicated interaction with the surrounding gas cloud.

The structure of our work is as follows. In Section~2, we describe the numerical setup of the simulations, the chemistry and cooling model, and our method for finding protostars and determining their properties. In Section~3, we present the simulations and analyse the initial collapse of the minihaloes, the formation and fragmentation of the disc, the evolution of the protostellar system as a whole, and the merger and survival rates of the secondary protostars formed in the disc. Finally, in Section~4 we summarize our results and draw conclusions. All distances are quoted in proper units, unless noted otherwise.

\section{Numerical methodology}

The initial conditions of the simulations are taken from \citet{greif11a}, which are cosmological simulations that employ multiple levels of refinement to follow the collapse of the central gas clouds in minihaloes. We briefly review these simulations and describe the additions made to the code to follow the collapse of the gas well into the optically thick regime at $n_{\rm H}\ga 10^{17}\,{\rm cm}^{-3}$. All of our simulations were performed with the moving-mesh code {\sc arepo}.

\subsection{The moving-mesh code AREPO}

{\sc arepo} is a finite-volume code that employs a Voronoi tesselation of space to compute hydrodynamic fluxes across cell boundaries \citep{springel10a}. For a given ensemble of points, a cell is constructed around each point that contains all regions of space that are closest to the point under consideration. Apart from small corrective velocities to ensure that the mesh remains regular, the mesh-generating points are advected with the underlying fluid velocity and become quasi-Lagrangian. This special property of {\sc arepo} allows for a computation of Galilean-invariant fluxes, which is one of the main differences from finite volume methods on a fixed mesh. Even though the construction of the mesh may appear to be very expensive, the computational speed of {\sc arepo} is in fact similar to that of the smoothed particle hydrodynamics (SPH) code {\sc gadget} \citep{syw01,springel05}, since both employ expensive neighbour searches that dominate the runtime. We note that {\sc arepo} may also be employed for pure DM simulations, for which it uses the same gravitational oct-tree as in {\sc gadget-3} \citep[last described in][]{springel05}, but with a number of optimisations to the domain decomposition.

\begin{table*}
\begin{center}
\caption{Simulation Parameters}
\begin{tabular}{lcccccccccc} \hline \hline
Simulation & Size & Particles & $M_{\rm dm}$ & $M_{\rm dm, ref}$ & $M_{\rm gas}$ & $\sigma_{8}$ & $M_{\rm vir}$ & $r_{\rm vir}$ & $\lambda$ & $z_{\rm coll}$ \\
& (kpc) & & $({\rm M}_\odot)$ & $({\rm M}_\odot)$ & $({\rm M}_\odot)$ & & $({\rm M}_\odot)$ & $({\rm pc})$ & & \\ \hline
MH1 & $500$ & $256^3$ & $272$ & $3.53$ & $0.72$ & $0.9$ & $3.0\times 10^{5}$ & $110$ & $0.055$ & $19.5$ \\
MH2 & $250$ & $128^3$ & $272$ & $3.53$ & $0.72$ & $1.2$ & $2.3\times 10^{5}$ & $94$ & $0.073$ & $20.9$ \\
MH3 & $500$ & $256^3$ & $272$ & $3.53$ & $0.72$ & $1.1$ & $3.1\times 10^{5}$ & $97$ & $0.044$ & $22.6$ \\
MH4 & $500$  & $256^3$ & $272$ & $3.53$ & $0.72$ & $1.3$ & $1.8\times 10^{5}$ & $58$ & $0.038 $ & $31.7$ \\ \hline
\multicolumn{11}{l}{} \\
\multicolumn{11}{l}{\parbox{14cm}{The comoving box sizes, initial DM particle numbers, DM masses, refined DM masses, gas masses, and normalizations used in the cosmological simulations, followed by the virial masses, virial radii, spin parameters, and collapse redshifts of the first minihaloes that form. The halo properties agree well with those found in previous studies \citep[e.g.][]{mba01,yoshida03a,gao07, on07}.}}\\
\end{tabular}
\end{center}
\end{table*}

\subsection{Dark matter simulations}

We first use {\sc arepo} to perform a series of DM simulations and determine the locations of the minihaloes that are later flagged for refinement. The employed boxes have a side length of $250$ and $500\,{\rm kpc}$ (comoving), and use $128^3$ and $256^3$ particles of mass $M_{\rm dm}\simeq 272\,{\rm M}_\odot$. We use a comoving gravitational softening length of $68\,{\rm pc}$ and imprint an initial fluctuation power spectrum for a $\Lambda$ cold dark matter ($\Lambda$CDM) cosmology at a starting redshift of $z=99$ with matter density $\Omega_{\rm m}=1-\Omega_\Lambda=0.27$, baryon density $\Omega_{\rm b}=0.046$, Hubble parameter $h=H_0/100\,{\rm km}\,{\rm s}^{-1}\,{\rm Mpc}^{-1}=0.71$ (where $H_0$ is the present Hubble expansion rate) and a spectral index $n_{\rm s}=0.96$ \citep[e.g.][]{komatsu09}. The linearly extrapolated present-day normalization $\sigma_8$ is varied with respect to the measured value of $0.81$ to represent rare overdense regions of the Universe that cannot be captured with the limited box sizes employed here. The DM simulations are evolved until the first minihalo with a virial mass exceeding $5\times 10^5\,{\rm M}_\odot$ collapses. The parameters of the simulations as well as the properties of the minihaloes are summarized in Table~1.

\subsection{Refined simulations}

After the location of a minihalo has been determined, we construct a high-resolution study of the halo by first identifying all particles within approximately two times the virial radius. These particles are then traced back to their initial conditions, yielding the Lagrangian region that formed the halo. To construct new initial conditions, each particle in this region is replaced by $64$ DM particles and $64$ mesh-generating points that generate the space-filling cells that are used for the hydrodynamic calculations. This corresponds to the minimum resolution that is necessary to adequately resolve the collapse of the gas in minihaloes \citep[for a resolution study, see][]{greif11b}. We note that we do not include relative streaming velocities between the DM and the gas, which influences the virialization of the minihaloes and might have a non-negligible impact on the later evolution of the gas cloud \citep{greif11b}.

Cells and DM particles outside of the high-resolution region are replaced by ever higher-mass particles and cells with increasing distance from the target halo, such that the resolution coarsens significantly far away from the target halo, yet the gravitational tidal field that influences its formation is still followed with high accuracy. This step reduces the total number of particles and cells in the refined initial conditions to $\simeq 2\times 10^6$. The simulation box is then centred on the target halo and reinitialized at $z=99$ based on the original realization of the density field, but augmented with additional small-scale power in the high-resolution region that can now be represented. The initial DM particle and cell masses in the high-resolution region before any further run-time refinement are $M_{\rm dm, ref}=(1-\Omega_{\rm b}/\Omega_{\rm m})M_{\rm dm}/64\simeq 3.53\,{\rm M}_\odot$ and $M_{\rm gas}=(\Omega_{\rm b}/\Omega_{\rm m})M_{\rm dm}/64\simeq 0.72\,{\rm M}_\odot$, respectively. We use a comoving gravitational softening length of $17\,{\rm pc}$ for the refined DM component.

\subsection{Extraction procedure}

The refined simulations are run until the first cell exceeds a density of $n_{\rm H}=10^9\,{\rm cm}^{-3}$, after which we extract the central $1\,{\rm pc}$ and reinitialize the simulations with reflective boundary conditions, which are better suited to deal with highly asymmetric boundaries. We further remove all DM particles, which has almost no effect on the evolution of the central gas cloud, since the gas is already well decoupled from the DM component \citep{greif11a}. These initial resimulations are then evolved until the density exceeds $n_{\rm H}=10^{19}\,{\rm cm}^{-3}$, after which we again extract the central $2000\,{\rm au}$ and employ reflective boundary conditions. The sound crossing time of the refined box exceeds the simulation time by at least two orders of magnitude, such that perturbations from the edges of the box do not affect our results.

\subsection{(De-)refinement strategy}

The collapse of the gas to ever smaller scales leads to a continuous decrease of the Jeans length, which must be accompanied by a continuous refinement of the gas to ensure that the Truelove criterion is fulfilled \citep{truelove98}. The latter may be expressed by the minimum number of cells that must be employed per Jeans length:
\begin{equation}
N_{\rm J, min}=4\frac{\lambda_{\rm J}}{h},
\end{equation}
where $\lambda_{\rm J}$ is the Jeans length and $h$ is the radius of a cell, which is defined using the volume of the cell: $h^3=3V/4\pi$. Strictly, the above criterion applies only to isothermal gas. However, it is commonly extended to gas with a variable equation of state by using a minimum of eight cells per Jeans length \citep[e.g.][]{turk12}. We employ a somewhat modified version of the Jeans refinement described in \citet{greif11a}. Specifically, we refine a cell once its density exceeds $n_{\rm H}=1\,{\rm cm}^{-3}$ within the central $200\,{\rm pc}$ of the halo and if, in addition, the Jeans length falls below the target number of cells per Jeans length, which is given by
\begin{equation}
N_{\rm J, t}=N_{\rm J, a}-(N_{\rm J, a}-N_{\rm J, b}) \frac{\log{\left(n_{\rm H}/n_{\rm H, a}\right)}}{\log{\left(n_{\rm H, b}/n_{\rm H, a}\right)}},
\end{equation}
where $n_{\rm H, a}=10^9\,{\rm cm}^{-3}$, $n_{\rm H, b}=10^{21}\,{\rm cm}^{-3}$, $N_{\rm J, a}=128$, and $N_{\rm J, b}=32$. For $n_{\rm H}<n_{\rm H, a}$, we keep $N_{\rm J, t}$ fixed at $N_{\rm J, a}$, and for $n_{\rm H}>n_{\rm H, b}$ we keep $N_{\rm J, t}$ fixed at $N_{\rm J, b}$. The utilisation of $128$ cells per Jeans length at low densities ensures that the turbulence generated during the initial collapse is represented adequately \citep[e.g.][]{federrath11,turk12}, while the lower resolution of $32$ cells per Jeans length at high densities ensures that the simulations remain computationally feasible. The maximum spatial resolution achieved with this refinement strategy is about $0.05\,{\rm R}_\odot$ at the center of the protostars, and gradually decreases to about $10\,{\rm au}$ at the outskirts of the box. We have found that a slight increase of $N_{\rm J, b}$ to $64$ results in a substantial increase of the necessary computing time, which cannot be easily compensated by using a larger number of processors. Each simulation was run for about three months on $32$ processors, and employed several million internal timesteps to evolve the simulations for about $10\,{\rm yr}$. The exact simulated timespan varied somewhat, since the physical differences between the minihaloes also led to differences in the timespan simulated per wall clock time.

The cells that are flagged for refinement are split into two cells by replacing the mesh generating point with a close pair of points. The conserved fluid quantities in the original cell are distributed on to the two halves in proportion to their volume fractions. In the subsequent evolution, the pair of mesh-generating points separates, leading to a smooth adjustment of the local mesh. We also use the derefinement procedure of {\sc arepo} to save computational time in regions where the collapse has stalled. If the size of a cell exceeds the Jeans length by a factor of two, and if in addition it is part of a smooth region of the flow (measured by requiring that the density, velocity, pressure, and temperature gradients across a cell are small), we remove the cell's mesh-generating point and distribute its content on to the neighbouring cells in a conservative fashion. The relative fraction given to each of the neighbouring cells is determined by the fraction of the volume that they claim from the removed cell, as determined by the Voronoi tessellation used in {\sc arepo}. We note that the chemical species are not redistributed in this fashion. The abundance information contained in the target cell is therefore discarded, which is a simplification that needs to addressed in future work.

\subsection{Chemistry and cooling}

We have substantially revised the chemical network employed in \citet{greif11a} to include a non-equilibrium solver at low densities and an equilibrium solver at high densities. The non-equilibrium solver uses the publicly available ordinary differential equation solver {\small SUNDIALS CVODE} \citep{hindmarsh05} to compute the temperature and abundances for a set of non-linearly coupled chemical and thermal rate equations. The evolved species consist of H, H$^+$, H$^-$, H$_2^+$, H$_2$, He, He$^+$, He$^{++}$, D, D$^+$, HD, and free electrons \citep{gj07,clark11a}. The thermal energy of the gas is evolved alongside the other rate equations. The most important coolant at low densities is molecular hydrogen, which is collisionally excited and decays radiatively via rovibrational transitions. A second, less important coolant is hydrogen deuteride (HD), which may become relevant at low temperatures due to its permanent dipole moment \citep{flower00,nu02,gb06,jb06,ripamonti07,mb08}. In fact, two of the minihaloes investigated here cool to $\simeq 100\,{\rm K}$ through HD line emission before becoming Jeans-unstable \citep[see also][]{greif11a}. As will be discussed in Section~3, their evolution differs somewhat from the haloes that have only cooled via H$_2$. We note that we neglect any reactions involving deuterium species above a density of $10^{10}\,{\rm cm}^{-3}$, since they no longer affect the chemical evolution of the gas.

Above densities of $10^8\,{\rm cm}^{-3}$, three-body reactions convert the remaining atomic hydrogen into a fully molecular gas. We derive the three-body formation rate by applying the principle of detailed balance to the collisional dissociation rate of H$_2$ quoted in \citet{pss83}. The resulting rate is intermediate in terms of the large uncertainty discussed in \citet{ga08}, which is reflected by the substantial variation of the thermal and morphological evolution of primordial gas clouds found in \citet{turk11}. For $n_{\rm H}\ga 10^9\,{\rm cm}^{-3}$, column densities are high enough that the gas becomes optically thick to molecular hydrogen lines, and we determine the escape fration using the Sobolev approximation. At densities $\ga 10^{14}\,{\rm cm}^{-3}$, collision-induced emission (CIE) becomes important and provides the last radiative cooling channel \citep{on98,ra04}. However, it is quickly suppressed by the continuum opacity of the gas \citep{ripamonti02,ra04}. Finally, collisional dissociation of H$_2$ removes $4.48\,{\rm eV}$ of thermal energy per molecule from the gas and substantially offsets the compressional heating up to very high densities \citep{omukai00,yoh08}.

Once the density exceeds $10^{14}\,{\rm cm}^{-3}$, the timestep of the non-equilibrium solver becomes prohibitively small, which is due to the dependence of the three-body reactions on the cube of the density. We therefore switch to an equilibrium solver on a cell-by-cell basis for $n_{\rm H}>n_{\rm thresh}=10^{14}\,{\rm cm}^{-3}$. We determine the H$_2$ fraction according to the ratio of the three-body formation and collisional destruction rates, which in our case is given by
\begin{equation}
\frac{n_{{\rm H}_2}}{(n_{\rm H}-2n_{{\rm H}_2})^2}=\frac{1.05\times 10^{-22}}{T^{0.515}}\exp{\left(\frac{5.2\times10^4\,{\rm K}}{T}\right)},
\end{equation}
where $n_{{\rm H}_2}$ is the number density of hydrogen molecules and $T$ the temperature of the gas. The H$^+$ fraction is computed in a similar fashion \citep[see also][]{yoh08}:
\begin{equation}
\frac{n^2_{{\rm H}^+}}{n_{\rm H}-n_{{\rm H}^+}}=\left(\frac{2\pi m_{\rm e}k_{\rm B}T}{h^2_{\rm P}}\right)^{3/2}\exp{\left(-\frac{\chi_{\rm H}}{k_{\rm B}T}\right)},
\end{equation}
where $n_{{\rm H}^+}$ is the number density of ionised hydrogen atoms, $m_{\rm e}$ the electron mass, $k_{\rm B}$ Boltzmann's constant, $h_{\rm P}$ Planck's constant and $\chi_{\rm H}$ the ionization energy of hydrogen. For simplicity, we assume that the He$^+$ abundance is equal to the H$^+$ abundance, multiplied by the ratio of helium to hydrogen nuclei, and set the He$^{++}$ abundance to zero. We do not expect our results to be sensitive to this simplification. We converge on the self-consistent temperature and H$_2$ fraction via iteration, which yields the chemical heating and cooling rate of the gas by H$_2$ formation and dissociation.

We note that we have neglected a number of additional cooling mechanisms that may become important at the high densities encountered in our simulations. Among these are H$^-$ formation cooling, Ly$\alpha$ cooling, and collisional ionization cooling. A more sophisticated treatment would also require full radiative transfer of the various line and continuum processes, instead of assuming a local escape fraction. An upper limit on the heating of the gas by accretion luminosity from the central protostar was investigated in \citet{clark11b}, \citet{greif11a}, and \citet{smith11a}, where the gas was assumed to be optically thin to blackbody emission from the protostar. In this case the fragmentation of the gas in the inner few au was suppressed. However, since we follow the collapse of the gas to very high densities, we refrain from using an optically thin treatment and instead use escape fractions.

Finally, we note that the chemistry solver employed here assumes that the electron fraction transitions into equilibrium at the same density as H$_2$, instead of at $n_{\rm H}\ga 10^{17}\,{\rm cm}^{-3}$, where the transition is expected to occur. We therefore underestimate the electron fraction once the gas exceeds the threshold density for the first time, but its abundance remains so low that it does not influence the initial collapse of the gas. This is no longer true once a significantly ionised parcel of gas expands and crosses the threshold density in the reverse direction, which leads to an unphysical jump in the electron fraction. We have verified that only a small fraction of the mass and volume of the gas at the outskirts of the cloud are subject to this problem, so that it hardly affects our results, but it should be addressed in future work.

\subsection{Protostar finder}

The further analysis requires an identification of protostars. Their extent is commonly determined by using the location of the accretion shock, but in the present study we use the photosphere as the boundary of the protostars, which is more straightforward to find computationally. Both approaches should yield similar results, since the opacity of the gas rises sharply at the accretion shock. The photosphere is determined with a post-processing algorithm that finds the spherically averaged radius at which the optical depth exceeds unity. In addition, we track the positions and properties of the protostars, which allows us to construct merger histories. The relevant calculations are performed on particle dumps (snapshots) that are output every $\simeq 0.02\,{\rm yr}$.

Starting from the first snapshot, we locate the densest cell in the simulation with at least $n_{\rm H}=10^{19}\,{\rm cm}^{-3}$. This cell is defined as the centre of a new protostar, provided that the distance to the centre of any other protostar exceeds the radius of the protostar as well as a predefined merger radius $r_{\rm merge}$, which we set to $0.1\,{\rm au}$. The above density threshold gives a good indication of when the collapse of the gas stalls, a shock forms, and a new protostar is created. Since the center of a new protostar must lie outside of all existing protostars, our results are not very sensitive to this parameter. The above choice for the merger radius reliably identifies when two protostars finally merge. We have found that a significantly smaller value may suppress mergers, since the centers of two protostars that merge do not necessarily move to the exact same location, and a significantly larger value may overproduce mergers, since protostars may overlap temporarily on scales well below $1\,{\rm au}$.

Once a candidate cell has been selected, we determine the optical depth of $N_{\rm ang}\simeq 10^4$ nearly uniformly spaced angular bins and $N_{\rm rad}=200$ logarithmically spaced radial bins between $0.01$ and $10\,{\rm au}$ centred on the candidate cell:
\begin{equation}
\Delta\tau_{j, k}=\rho_{j, k}\kappa_{j, k}\Delta r_k,
\end{equation}
where $j$ denotes the angular bin, $k$ the radial bin, $\rho_{j, k}$ the mass-weighted density of the bin, $\kappa_{j, k}$ the Rosseland mean opacity, and $\Delta r_k$ the radial extent of the bin. For primordial gas in chemical and thermal equilibrium, which is a good approximation since the density typically exceeds $n_{\rm H}=10^{17}\,{\rm cm}^{-3}$ in the protostars, the opacity is only a function of density and temperature, and has been tabulated by \citet{md05}. We linearly interpolate from their tables to determine the mean opacity of each bin from the mass-weighted density and temperature of all cells contributing to the bin.

The integrated optical depth is obtained from equation~(5) by summing up the differential optical depths from large to small radii:
\begin{equation}
\tau_{j, k}=\sum_{l=N_{\rm rad}-1}^{l=k}\Delta\tau_{j, l},
\end{equation}
We then find the radial index $k_{\rm crit}$ at which the spherically averaged escape fraction, given by
\begin{equation}
\beta_{{\rm esc}, k}=\frac{1}{N_{\rm ang}}\sum_j\frac{1-\exp{\left(-\tau_{j, k}\right)}}{\tau_{j, k}}
\end{equation}
drops to $\beta_{\rm crit}=1-\exp{\left(-1\right)}\simeq0.63$, which corresponds to an optical depth of unity. The photospheric radius of the protostar $R_*$ is then simply identified with $r_k(k_{\rm crit})$, and the mass of the protostar $M_*$ is given by the mass enclosed within $R_*$. Note that this definition implies that individual protostars may overlap and contain the same cells, which typically results in similar protostellar masses just before a merger occurs.

Once a protostar has been created, it is assigned a unique ID, and in addition we save the ID's of all cells that lie within the merger radius of the protostar. The position of the protostar in the next timestep is then determined from the new centre of mass of the saved cells. If this position lies within the merger radius of another protostar, it is merged with that protostar. The sequence of ID's therefore also tracks the formation history of the protostars. After the merging is complete, we continue with the first step and again find new protostars. This procedure is repeated until no candidate cells remain and all protostars in the snapshot have been identified. Finally, we loop over all snapshots until the entire history of the protostellar system has been reconstructed.

\section{Results}

\subsection{Collapse of central gas cloud}

Many aspects of the collapse of primordial gas in minihaloes have been discussed in previous work \citep[e.g.][]{abn02,bcl02,yoshida03a,yoshida06b,yoh08}. We expand on these studies by comparing the baryonic properties of the minihaloes investigated here over an unprecedented range in scales. In Fig.~1, we show a number of physical quantities as a function of the distance to the densest cell in each minihalo. The haloes are denoted by different line styles, which have been constructed using data from two snapshots: the inner $10^5\,{\rm au}$ are extracted from the final snapshots of the initial resimulations, when the density of hydrogen nuclei first exceeds $10^{19}\,{\rm cm}^{-3}$, and the outer $10^5-10^7\,{\rm au}$ are taken from the final snapshot of the original cosmological simulations, when the density of hydrogen nuclei first exceeds $10^{9}\,{\rm cm}^{-3}$.  The maximum density of $10^{19}\,{\rm cm}^{-3}$ gives a good estimate of when the collapse of the gas stalls and a shock forms at the protostellar surface. Due to the self-similar nature of the collapse, moving from large to small radii may also be viewed as a progression in time. Fig.~1 therefore also shows the evolution of the central gas cloud from the assembly of the halo until the formation of the first protostar.

From top left to bottom right, the panels in Fig.~1 show the density of hydrogen nuclei, enclosed gas mass, temperature, H$_2$ fraction, e$^-$ fraction, radial velocity, sound speed, radial Mach number, rotation velocity, ratio of rotation to Keplerian velocity, angle between the vector of the rotation velocity of a given radial bin and the innermost radial bin, and turbulent Mach number. With a few exceptions, these quantities are determined by using the mass-weighted average of the cells contributing to each bin. In panels involving the radial or rotation velocity, we first subtract the bulk velocity of the halo from each cell. We determine the rotation velocity by adding up the angular momentum of all contributing cells, and divide it by the mass and radius of the bin. The angle $\phi$ is given by the angle between the vector of the rotation velocity of a given radial bin and the innermost bin. We define the turbulent Mach number as
\begin{equation}
M_{\rm turb}^2c_{\rm s}^2=\sum_i \frac{m_i}{M}\left({\bmath v}_i-{\bmath v}_{\rm rad}-{\bmath v}_{\rm rot}\right)^2,
\end{equation}
where $c_{\rm s}$ denotes the sound speed of the radial bin, $M$ the total mass, ${\bmath v}_{\rm rad}$ the vector of the radial velocity, ${\bmath v}_{\rm rot}$ the vector of the rotation velocity, $i$ the index of a cell contributing to the bin, $m_i$ its mass, and ${\bmath v}_i$ its velocity.

To first order, the collapse of the gas follows a Larson-Penston solution for an isothermal, self-gravitating gas cloud \citep{larson69,penston69}. However, since the temperature of the gas varies and the cloud possesses some degree of turbulent and rotational support, the radial density profiles deviate somewhat from the idealised $r^{-2}$ profile. The scatter in densities at a given radius can exceed an order of magnitude, showing that a significant amount of substructure is present. The virialisation heating of the gas as it falls into the DM halo can be seen at about $10^{7}\,{\rm au}$ ($\simeq 50\,{\rm pc}$) in the temperature profile. As the density and temperature rise, the molecular hydrogen fraction builds up to about $10^{-3}$ and the gas cools to about $200\,{\rm K}$ on a scale of $\simeq 10^5\,{\rm au}$ ($\simeq 1\,{\rm pc}$). During this initial free-fall phase, the turbulent Mach number typically exceeds unity. When the H$_2$ level populations transition into equilibrium and the cooling rate no longer scales with the density squared, the gas begins to heat up rapidly to about $1000\,{\rm K}$, and the turbulent Mach number decreases again. Note that in simulations MH1 and MH2, HD cooling becomes relevant and the gas cools to about $100\,{\rm K}$. Although the temperature offset has nearly disappeared by the time the gas has collapsed to about $10\,{\rm au}$, the radial velocity in these haloes is about a factor of two lower than in the other haloes, and they have also built up less turbulence.

On a scale of about $10^3\,{\rm au}$, three-body reactions rapidly convert the atomic hydrogen into a fully molecular gas. The resulting heating is relatively mild, since the growing H$_2$ abundance also ensures additional cooling. Once the gas becomes optically thick to H$_2$ cooling on a scale of about $100\,{\rm au}$, the temperature profile steepens again, and the gas heats up to a few thousand Kelvin. The inner $\simeq 1\,{\rm au}$ then begin to cool via CIE and H$_2$ collisional dissociation, which mitigates the rise in temperature as the gas further contracts. After most of the H$_2$ has been destroyed, the gas can no longer cool and heats up nearly adiabatically to almost $10^4\,{\rm K}$, resulting in a rapid increase of the electron fraction.

The rotation velocity remains comparable to the infall velocity throughout the entire evolution of the gas cloud, and its ratio to the Keplerian velocity varies between about $0.2$ and $0.7$, indicating a high degree of rotational support. The peak of the profile is at $\simeq 10^4\,{\rm au}$ after a period of strong heating, when the ratio of cooling time to dynamical time is large and the gas has time to settle into a disc. The rotation angle varies significantly during the collapse of the gas, which indicates vigorous transport of angular momentum. The main agent for angular momentum transport during the initial collapse is turbulence, which is comparable in magnitude to the Mach number of the radial infall, and typically of the order of unity.

We note that the properties of the gas in the initial collapse phase agree well with previous work. For example, the minimum temperature of about $200\,{\rm K}$ in our simulations is typically located at about $1\,{\rm pc}$, which is nearly identical to where the minimum is located in \citet{abn02} and \citet{on07}. The subsequent rise of the temperature to about $1000\,{\rm K}$ also occurs at nearly identical radii. The ratio of the rotation velocity to the Keplerian velocity in \citet{abn02} is about $0.3$, which is bracketed by the values we find here of between $0.2$ and $0.7$. At radii larger than about $10\,{\rm au}$, the radial velocity of the gas varies between about $1$ and $6\,{\rm km}\,{\rm s}^{-1}$, and is similar to what was found in \citet{on07}, even though we evaluate the gas properties at a later point in time and use a somewhat more comprehensive chemical network. At high densities, we find good qualitative agreement with \citet{yoh08}, which is the only study next to ours that evolves the collapse into the optically thick regime. In both cases, the electron fraction rises and the H$_2$ fraction drops significantly at about $0.1\,{\rm au}$, accompanied by an increase of the temperature to about $10^4\,{\rm K}$. The exact values have a relatively large scatter around those found in \citet{yoh08}, where only a single minihalo was investigated.

\begin{figure*}
\begin{center}
\includegraphics[width=16cm]{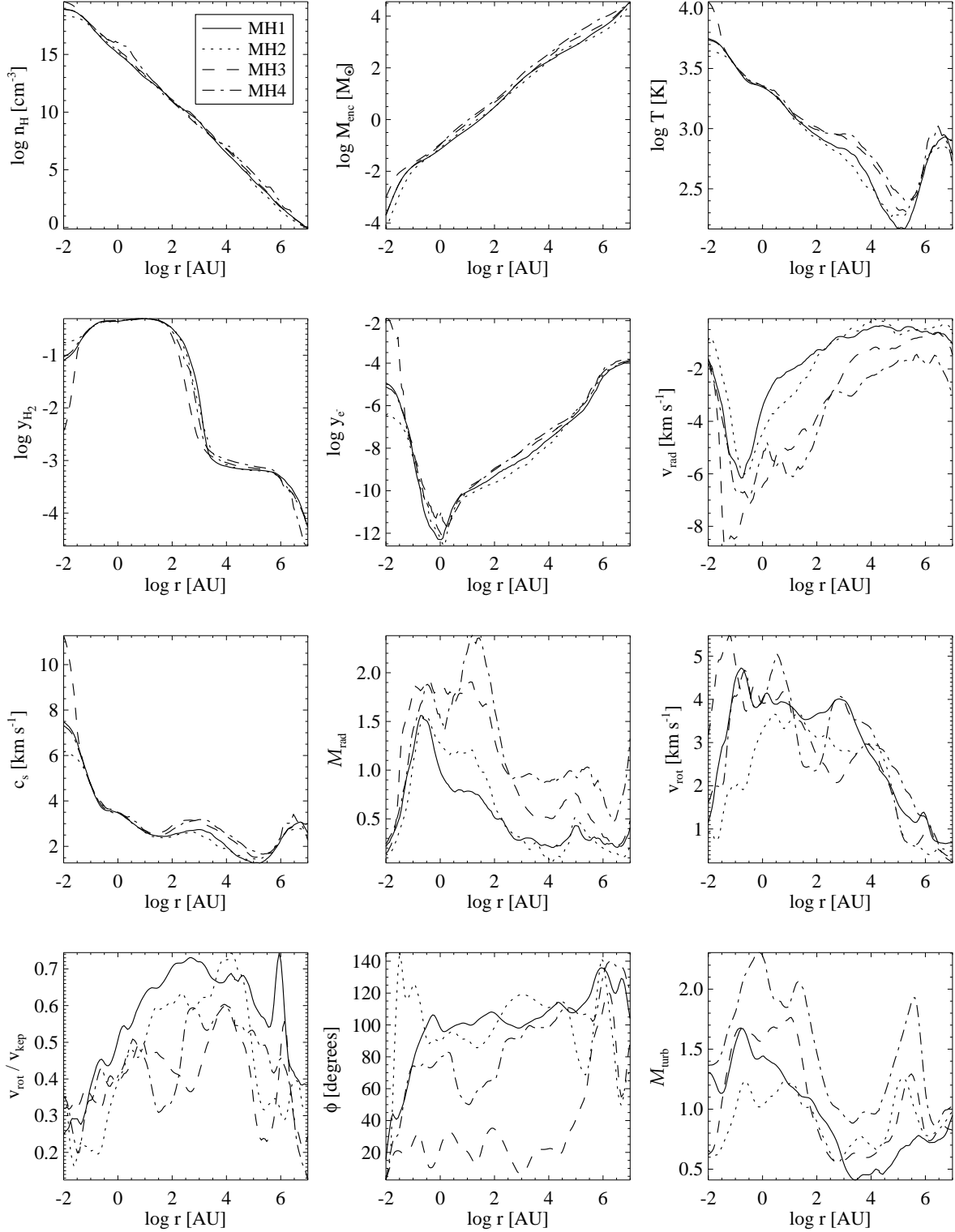}
\caption{From top left to bottom right: density of hydrogen nuclei, enclosed gas mass, temperature, H$_2$ fraction, e$^-$ fraction, radial velocity, sound speed, radial Mach number, rotation velocity, ratio of rotation to Keplerian velocity, angle between the vector of the rotation velocity at a given radial bin and the innermost radial bin, and turbulent Mach number versus distance to the densest cell in the simulation. The various line styles denote the minihaloes according to the legend. The above quantities are calculated when the density first exceeds $n_{\rm H}=10^{19}\,{\rm cm}^{-3}$ and a shock forms at the surface of the primary protostar, which is located at about $0.01\,{\rm au}$. The plot covers all relevant length scales and shows the collapse of the gas from cosmological to protostellar scales (see Section~3.1 for a detailed analysis).}
\end{center}
\end{figure*}

\begin{figure}
\begin{center}
\includegraphics[width=8cm]{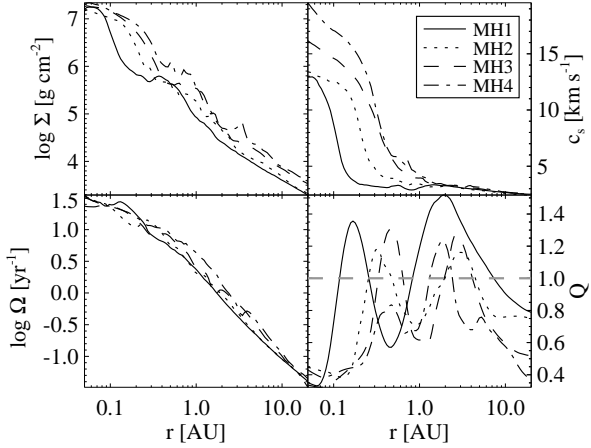}
\caption{From top left to bottom right: surface density, sound speed, rotation rate and Toomre $Q$ parameter versus distance to the centre of the primary protostar. The various line styles denote the minihaloes according to the legend. The above quantities are calculated just before the disc fragments. The growth of perturbations requires a Toomre $Q$ parameter below unity. This is the case below about $0.1\,{\rm au}$ and around $1\,{\rm au}$, which leads to the formation of spiral arms in the disc.}
\end{center}
\end{figure}

\begin{figure}
\begin{center}
\includegraphics[width=8cm]{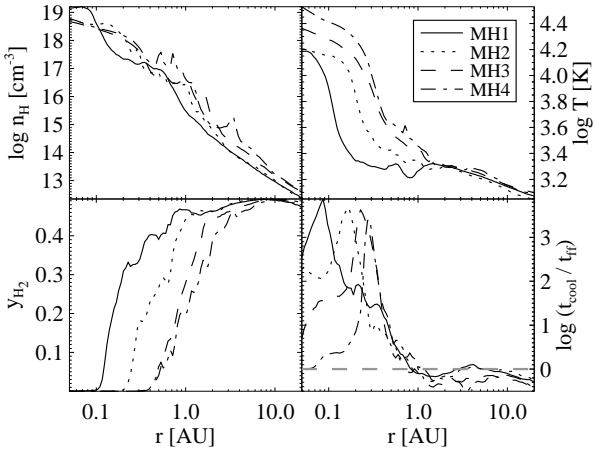}
\caption{From top left to bottom right: density of hydrogen nuclei, temperature, H$_2$ fraction and ratio of cooling time to the free-fall time versus distance to the centre of the primary protostar. The various line styles denote the minihaloes according to the legend. The above quantities are calculated just before the disc fragments. The temperature of the gas rises only moderately as long as sufficient H$_2$ is present and dissociation cooling is active. Below about $1\,{\rm au}$, cooling becomes very inefficient and the ratio of the cooling time to the free-fall time increases dramatically. Fragmentation occurs on a scale of about $1\,{\rm au}$, where both the $Q$ parameter and the time-scale ratio drop below unity.}
\end{center}
\end{figure}

\begin{figure}
\begin{center}
\includegraphics[width=8cm]{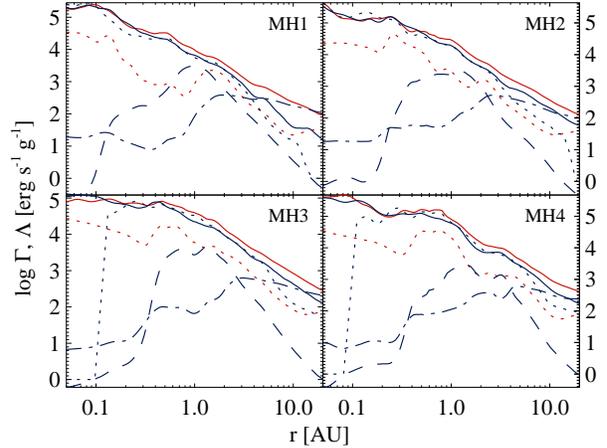}
\caption{Heating and cooling processes that operate in the disc just before it fragments. The various line styles denote compressional heating (red solid line), expansion cooling (blue solid line), H$_2$ formation heating (red dotted line), H$_2$ collisional dissociation cooling (blue dotted line), collision-induced emission (blue dashed line) and H$_2$ line cooling (blue dot-dashed line). Each panel shows a different minihalo. The most important heating process is compressional heating, which is counteracted by expansion cooling and H$_2$ collisional dissociation cooling. Collision-induced emission peaks on a scale of about $1\,{\rm au}$, while H$_2$ line cooling ceases to be important below about $10\,{\rm au}$.}
\end{center}
\end{figure}

\subsection{Disc formation and fragmentation}

After the first protostar has formed, the gas becomes fully rotationally supported in a Keplerian disc. The stability of the disc to perturbations may be quantified with the Toomre $Q$ parameter \citep{toomre64}:
\begin{equation}
Q=\frac{c_{\rm s}\kappa}{\pi G\Sigma},
\end{equation}
where $\kappa$ is the epicyclic frequency of the disc, $\Sigma$ the surface density, and $c_{\rm s}$ the sound speed of the gas. We replace the epicyclic frequency $\kappa$ with the orbital frequency $\Omega$, which is appropriate for Keplerian discs. The $Q$ parameter determines whether perturbations in an infinitely thin, isothermal disc can grow. For thick discs, as is the case here, a similar criterion may be found that deviates only by a factor of of the order of unity \citep[e.g.][]{wang10}. For $Q\ga 1$, the pressure of the gas and the shear by the differential rotation of the disc are sufficient to prevent local collapse, while for $Q\la 1$ the disc fulfils the Toomre criterion and becomes unstable, leading to the formation of spiral arms that transport mass inwards and angular momentum outwards.

\begin{figure*}
\begin{center}
\resizebox{16.2cm}{12.3cm}
{\unitlength1cm
\begin{picture}(16.2,12.3)
\put(0,9.6){\includegraphics[width=2.7cm,height=2.7cm]{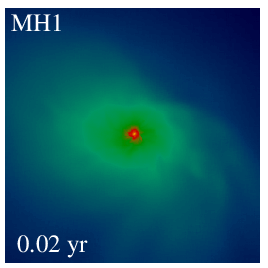}}
\put(2.7,9.6){\includegraphics[width=2.7cm,height=2.7cm]{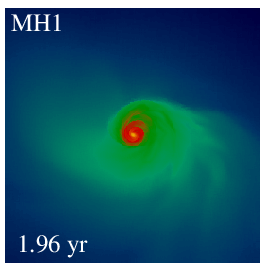}}
\put(5.4,9.6){\includegraphics[width=2.7cm,height=2.7cm]{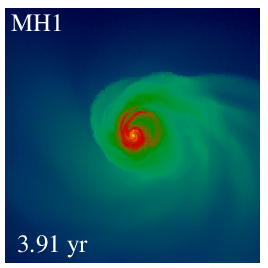}}
\put(8.1,9.6){\includegraphics[width=2.7cm,height=2.7cm]{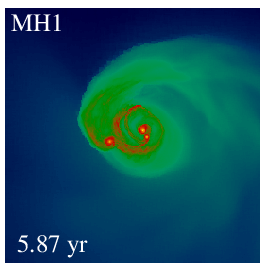}}
\put(10.8,9.6){\includegraphics[width=2.7cm,height=2.7cm]{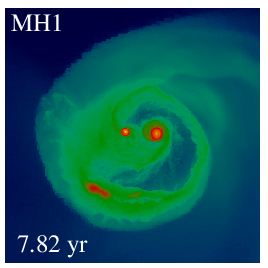}}
\put(13.5,9.6){\includegraphics[width=2.7cm,height=2.7cm]{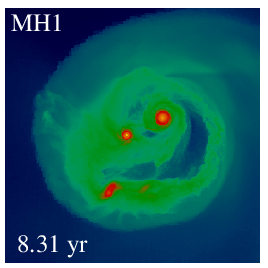}}
\put(0,6.9){\includegraphics[width=2.7cm,height=2.7cm]{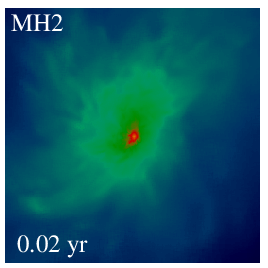}}
\put(2.7,6.9){\includegraphics[width=2.7cm,height=2.7cm]{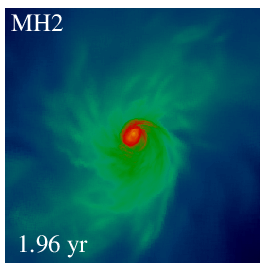}}
\put(5.4,6.9){\includegraphics[width=2.7cm,height=2.7cm]{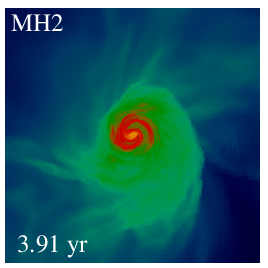}}
\put(8.1,6.9){\includegraphics[width=2.7cm,height=2.7cm]{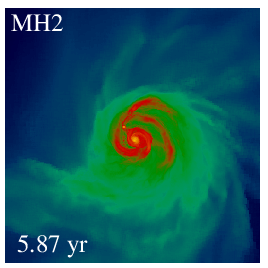}}
\put(10.8,6.9){\includegraphics[width=2.7cm,height=2.7cm]{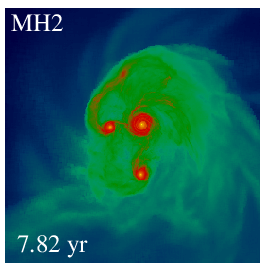}}
\put(13.5,6.9){\includegraphics[width=2.7cm,height=2.7cm]{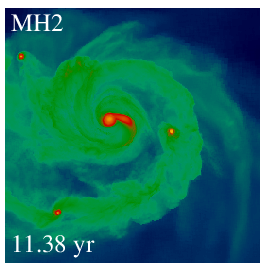}}
\put(0,4.2){\includegraphics[width=2.7cm,height=2.7cm]{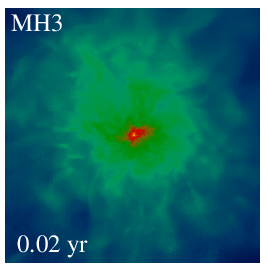}}
\put(2.7,4.2){\includegraphics[width=2.7cm,height=2.7cm]{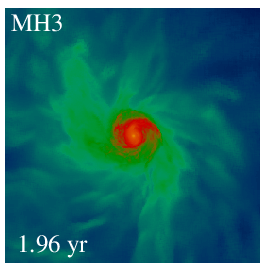}}
\put(5.4,4.2){\includegraphics[width=2.7cm,height=2.7cm]{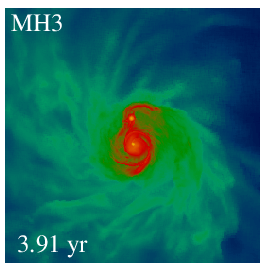}}
\put(8.1,4.2){\includegraphics[width=2.7cm,height=2.7cm]{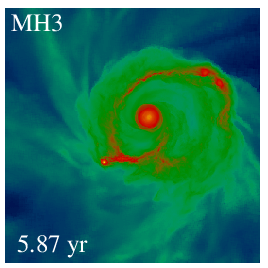}}
\put(10.8,4.2){\includegraphics[width=2.7cm,height=2.7cm]{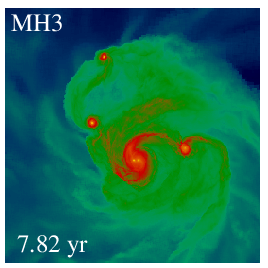}}
\put(13.5,4.2){\includegraphics[width=2.7cm,height=2.7cm]{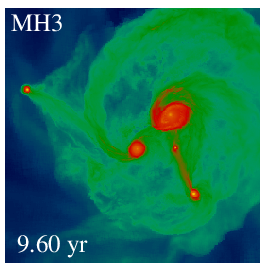}}
\put(0,1.5){\includegraphics[width=2.7cm,height=2.7cm]{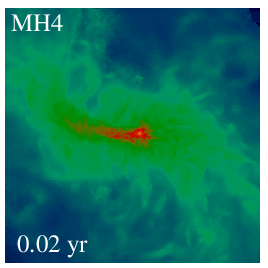}}
\put(2.7,1.5){\includegraphics[width=2.7cm,height=2.7cm]{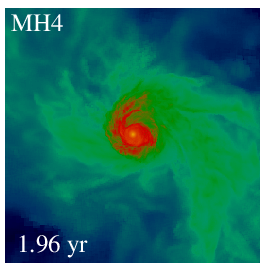}}
\put(5.4,1.5){\includegraphics[width=2.7cm,height=2.7cm]{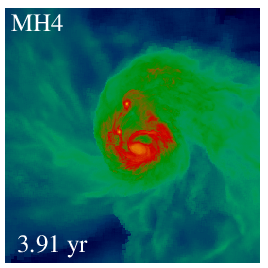}}
\put(8.1,1.5){\includegraphics[width=2.7cm,height=2.7cm]{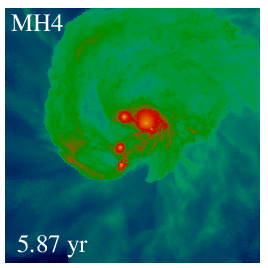}}
\put(10.8,1.5){\includegraphics[width=2.7cm,height=2.7cm]{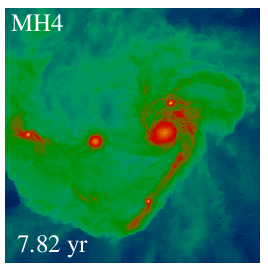}}
\put(13.5,1.5){\includegraphics[width=2.7cm,height=2.7cm]{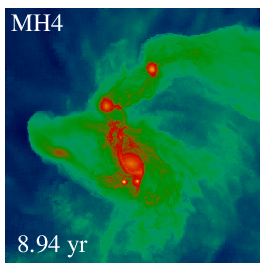}}
\put(2.1,0){\includegraphics[width=12cm,height=1.5cm]{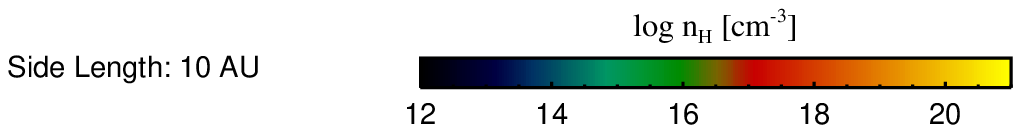}}
\end{picture}}
\caption{Density projections in a cube of side length $10\,{\rm au}$ that show the evolution of the protostellar system. Each row corresponds to a different minihalo. The time after the formation of the primary protostar increases from left to right. The final times vary since the physical differences between the minihaloes also result in different runtimes. The density of hydrogen nuclei is weighted by the density squared along the line of sight, which lies perpendicular to the plane of the disc. The disc around the primary protostar fragments into a number of secondary protostars, most of which migrate towards the centre of the cloud. However, some also obtain angular momentum from other protostars during close encounters and migrate to higher orbits. An example is the leftmost protostar at the last output time in MH3. A more detailed analysis of this figure is presented in Section~3.3.}
\end{center}
\end{figure*}

\begin{figure*}
\begin{center}
\resizebox{16.2cm}{12.3cm}
{\unitlength1cm
\begin{picture}(16.2,12.3)
\put(0,9.6){\includegraphics[width=2.7cm,height=2.7cm]{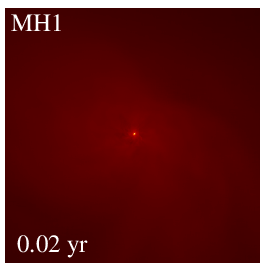}}
\put(2.7,9.6){\includegraphics[width=2.7cm,height=2.7cm]{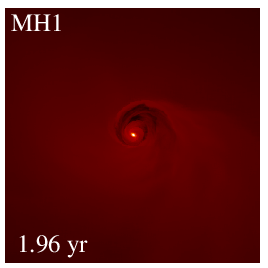}}
\put(5.4,9.6){\includegraphics[width=2.7cm,height=2.7cm]{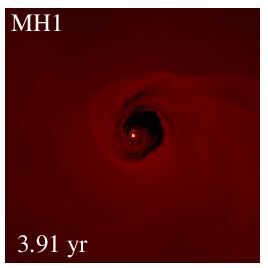}}
\put(8.1,9.6){\includegraphics[width=2.7cm,height=2.7cm]{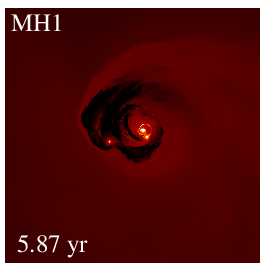}}
\put(10.8,9.6){\includegraphics[width=2.7cm,height=2.7cm]{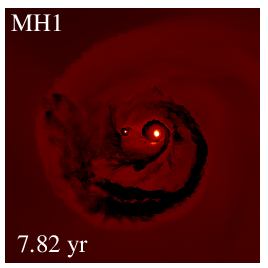}}
\put(13.5,9.6){\includegraphics[width=2.7cm,height=2.7cm]{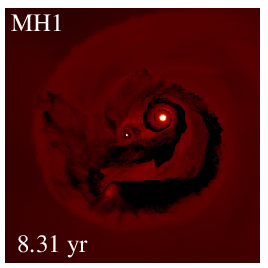}}
\put(0,6.9){\includegraphics[width=2.7cm,height=2.7cm]{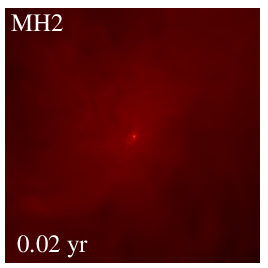}}
\put(2.7,6.9){\includegraphics[width=2.7cm,height=2.7cm]{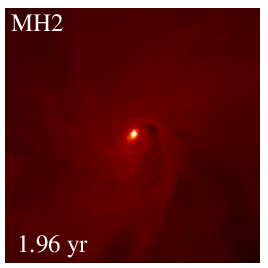}}
\put(5.4,6.9){\includegraphics[width=2.7cm,height=2.7cm]{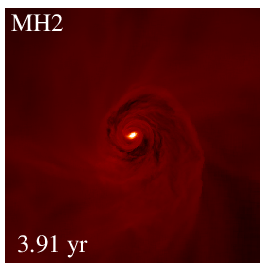}}
\put(8.1,6.9){\includegraphics[width=2.7cm,height=2.7cm]{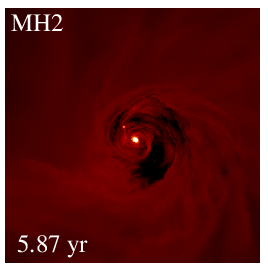}}
\put(10.8,6.9){\includegraphics[width=2.7cm,height=2.7cm]{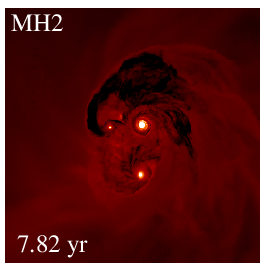}}
\put(13.5,6.9){\includegraphics[width=2.7cm,height=2.7cm]{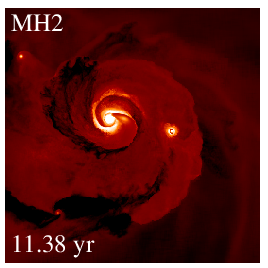}}
\put(0,4.2){\includegraphics[width=2.7cm,height=2.7cm]{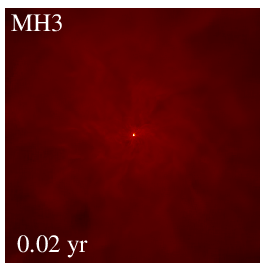}}
\put(2.7,4.2){\includegraphics[width=2.7cm,height=2.7cm]{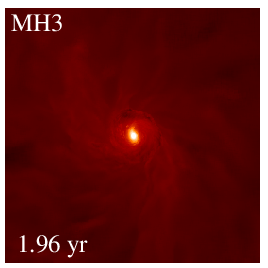}}
\put(5.4,4.2){\includegraphics[width=2.7cm,height=2.7cm]{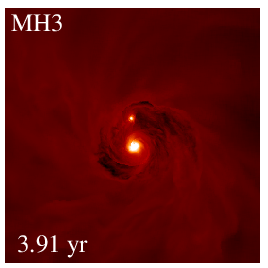}}
\put(8.1,4.2){\includegraphics[width=2.7cm,height=2.7cm]{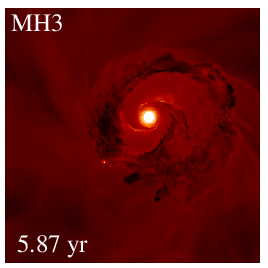}}
\put(10.8,4.2){\includegraphics[width=2.7cm,height=2.7cm]{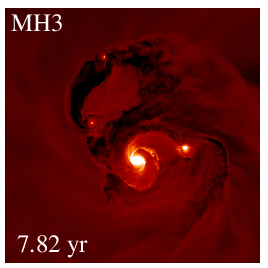}}
\put(13.5,4.2){\includegraphics[width=2.7cm,height=2.7cm]{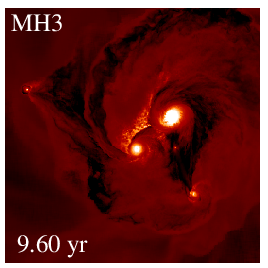}}
\put(0,1.5){\includegraphics[width=2.7cm,height=2.7cm]{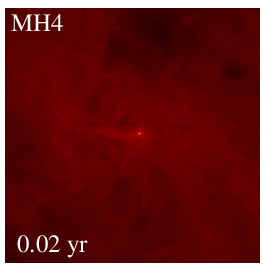}}
\put(2.7,1.5){\includegraphics[width=2.7cm,height=2.7cm]{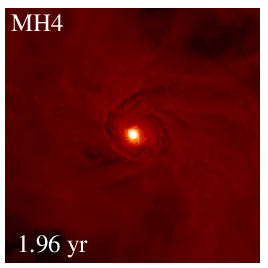}}
\put(5.4,1.5){\includegraphics[width=2.7cm,height=2.7cm]{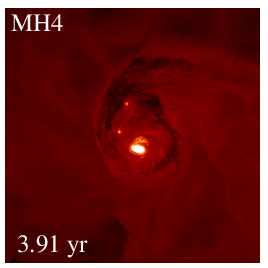}}
\put(8.1,1.5){\includegraphics[width=2.7cm,height=2.7cm]{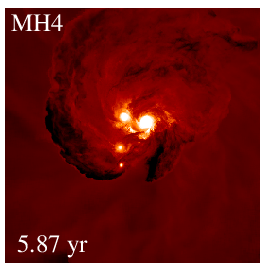}}
\put(10.8,1.5){\includegraphics[width=2.7cm,height=2.7cm]{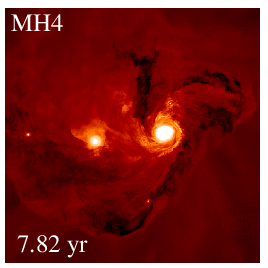}}
\put(13.5,1.5){\includegraphics[width=2.7cm,height=2.7cm]{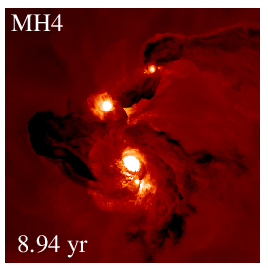}}
\put(2.1,0){\includegraphics[width=12cm,height=1.5cm]{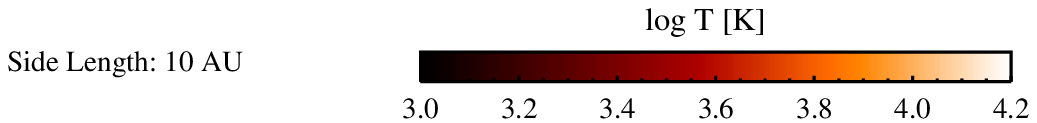}}
\end{picture}}
\caption{Temperature projections in a cube of side length $10\,{\rm au}$ showing the evolution of the protostellar system. Each row corresponds to a different minihalo. The time after the formation of the primary protostar increases from left to right. The temperature is weighted by the density squared along the line of sight, which lies perpendicular to the plane of the disc. The gas that accretes on to protostars is shock-heated to about $10^4\,{\rm K}$. From the extent of these regions it is evident that the primary protostar is generally much larger and more massive than the secondary protostars. The shock-heating of the gas ahead of spiral arms and the expansion cooling behind spiral arms are also apparent.}
\end{center}
\end{figure*}

In Fig.~2, we show the surface density, sound speed, orbital frequency, and $Q$ parameter in mass-weighted spherical shells around the densest cell in each minihalo just before the first fragment forms. We note that applying the same analysis exclusively to cells in the disc gives very similar results, since the mass within radial shells is dominated by the disc component. Within the central few au, the surface density is up to four orders of magnitude higher than in studies that have employed sink particles, and does not show a central dip around the sink particles \citep[e.g.][]{clark11b}. This is not surprising, since these studies typically do not resolve the gas on the scale of the accretion radius, and instead the mass is accreted on to the sink particles. The $Q$ parameter therefore does not diverge at small radii, but decreases to well below unity, which is reflected by the development of pronounced spiral arm patterns on sub-au scales at very early times. The orbital frequency on these scales drops to well below $10\,{\rm yr}^{-1}$. The rapid decline of the sound speed leads to a second minimum at $\simeq 1\,{\rm au}$, followed by an increase at larger radii due to the flat surface density profile and nearly constant sound speed. Interestingly, the gas does not fragment at the location of the first minimum, but at the second minimum at about $1\,{\rm au}$. This shows that $Q\la 1$ is not a sufficient criterion for gravitational instability and fragmentation.

A commonly employed criterion in addition to the Toomre criterion to evaluate whether the gas fragments is the Gammie criterion \citep{gammie01}:
\begin{equation}
t_{\rm cool}<\frac{3}{2\pi}t_{\rm orb},
\end{equation}
where $t_{\rm cool}$ is the cooling time and $t_{\rm orb}$ is the orbital time. Again, this equation strictly applies only to infinitely thin discs, but has also been shown to predict the fragmentation of thick discs quite well \citep{rice03}. In Fig.~3, we show the density of hydrogen nuclei, temperature, H$_2$ fraction, and ratio of cooling time to free-fall time using the same output times and binning as in Fig.~2. Note that the orbital time may be approximated by the free-fall time, since they differ only by a factor of the order of unity in a Keplerian disc. Moving from large to small radii, the ratio of cooling time to free-fall time rises sharply above unity once most of the H$_2$ has been dissociated and the gas can no longer cool, which is accompanied by a rapid rise in temperature as the gas heats up to $\ga 10^4\,{\rm K}$. This occurs on a scale of about $1\,{\rm au}$, which corresponds to the second minimum in the $Q$ parameter. This is the characteristic radius at which the disc fragments.

In Fig.~4, we investigate the various heating and cooling processes that result in the nearly isothermal radial profile of the disc down to very small radii. The rates are computed per unit mass instead of unit volume due to the strong increase in density towards the centre. We use the same output times and binning as in Figs~2 and 3. The individual lines correspond to compressional heating, expansion cooling, three-body H$_2$ formation heating, H$_2$ collisional dissociation cooling, CIE cooling, and H$_2$ line cooling. The most important heating process is compressional heating, which typically dominates over H$_2$ formation heating by an order of magnitude. The cooling of the gas is dominated by expansion cooling and collisional dissociation of H$_2$, which nearly balance the compressional heating rate. The CIE cooling curve peaks on a scale of about $1\,{\rm au}$, but otherwise has a relatively small contribution to the total cooling rate. H$_2$ line cooling is important on scales $\ga 10\,{\rm au}$, but is quickly diminished by the opacity of the gas on smaller scales. In MH3 and MH4, the dissociation cooling rate drops sharply at $\simeq 0.1\,{\rm au}$, since the molecular hydrogen reservoir is exhausted on larger scales as compared to MH1 and MH2 (see Fig.~3).

The above analysis shows that even though perturbations in the disc grow on scales as small as $\la 0.1\,{\rm au}$, the disc does not fragment on these scales since the gas is optically thick and the cooling time is larger than the dynamical time. Instead, fragmentation occurs on a scale of about $1\,{\rm au}$, where H$_2$ collisional dissociation and collision-induced emission cooling are sufficient to reduce the cooling time below the dynamical time. This is nearly an order of magnitude below the scale on which the first fragments formed in \citet{clark11b}, where the gas was assumed to be optically thin to radiation from the primary protostar and fragmentation was suppressed below about $10\,{\rm au}$. In \citet{greif11a}, an indication for a reduced fragmentation scale was presented in section~3.5, where in addition to the reduced fragment masses shown in fig.~16, the distance of the first fragment to the primary decreased as the accretion radius was reduced.

\begin{figure*}
\begin{center}
\resizebox{16cm}{21cm}
{\unitlength1cm
\begin{picture}(16,21)
\put(0,0){\includegraphics[width=8cm,height=21cm]{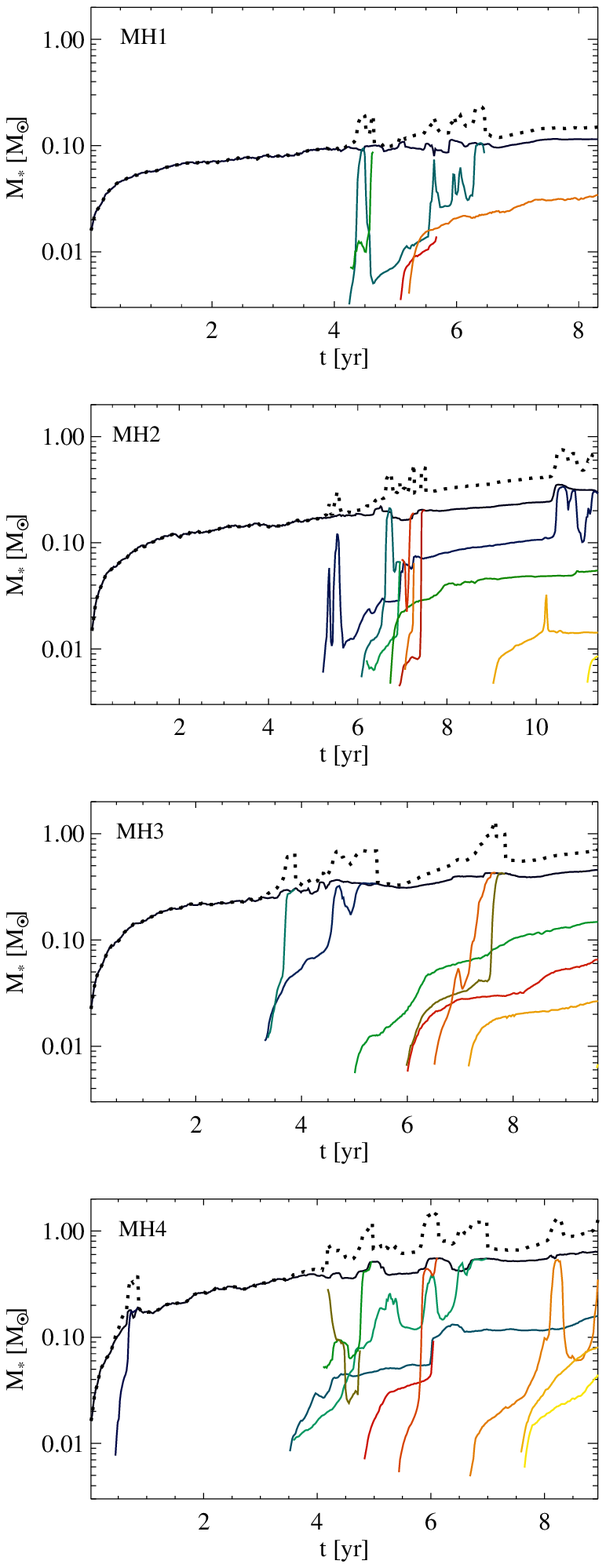}}
\put(8,0){\includegraphics[width=8cm,height=21cm]{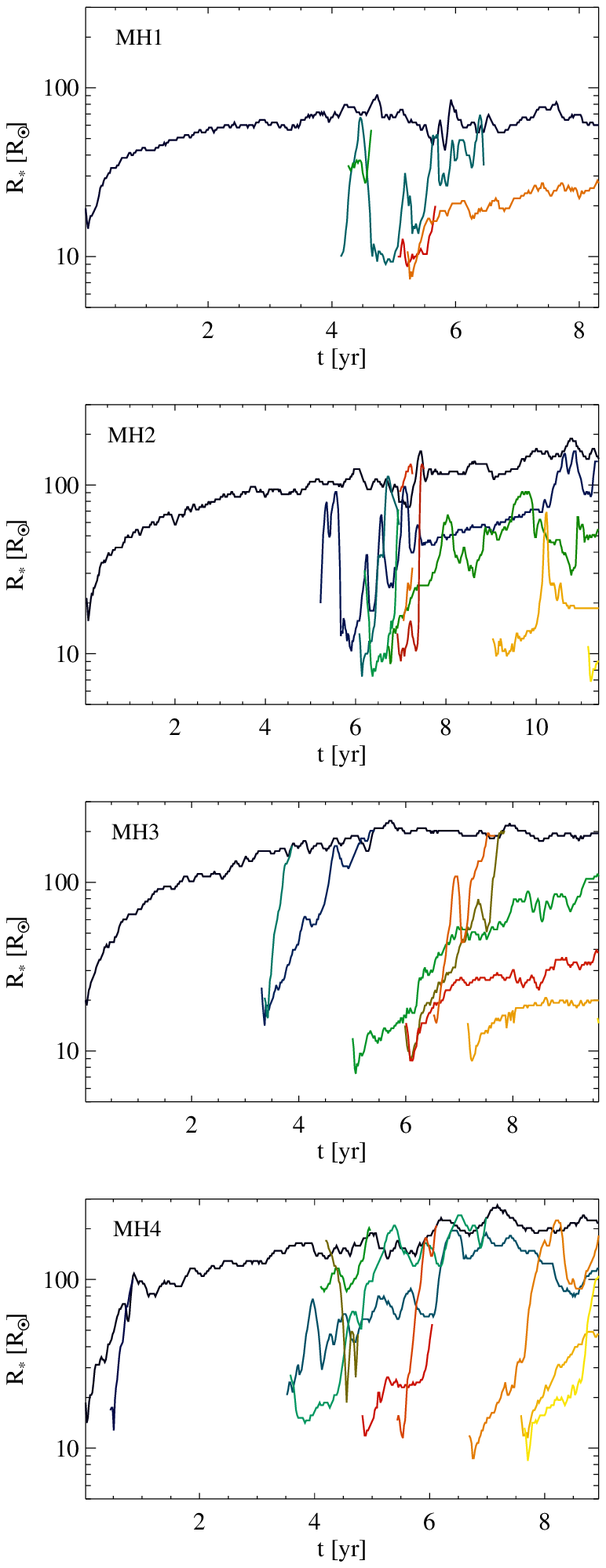}}
\end{picture}}
\caption{Left-hand panel: protostellar mass as a function of time after the formation of the primary protostar. Each line denotes an individual protostar, and the thick dotted lines denote the total mass of the protostellar system. Mergers are characterized by the equalization of two lines before one disappears. This aspect of the merging process was not captured by studies that employed sink particles, since protostars tend to spatially overlap for some time before they merge. The figure shows that merging is highly efficient and typically occurs between secondary protostars and the primary protostar. This facilitates the growth of the primary protostar to typically five times the mass of the second most massive protostar, and limits the number of secondary protostars that survive. Right-hand panel: Protostellar radius versus time. The radii of the protostars are in the expected range of $10-200\,{\rm R}_\odot$, with the primary protostar typically at least twice as large as the secondary protostars.}
\end{center}
\end{figure*}

\subsection{Evolution of protostellar system}

The fragmentation of the disc into one or more secondary protostars further complicates the evolution of the central gas cloud. This is evident from Fig.~5, where we show the number density of hydrogen nuclei in cubes with a side length of $10\,{\rm au}$ at six different output times. The line of sight is aligned perpendicular to the plane of the disc. The initial burst of star formation after the disc becomes gravitationally unstable is followed by a complicated succession of mergers between existing protostars and the formation of new protostars. By the end of the simulation, a small system of protostars ranging from a binary in MH1 to a system of five protostars in MH3 and MH4 have formed.

The temperature of the gas is shown in Fig.~6. The hottest parcels of gas reside within protostars and have temperatures in excess of $10^4\,{\rm K}$. The more massive protostars, and in particular the primary protostar, tend to reside at the centre of the cloud and are hotter than their low-mass counterparts at the outskirts of the cloud. The compressional heating of the gas ahead of spiral arms and the expansion cooling behind spiral arms are also apparent. Some striking differences between haloes that have activated HD cooling and those that have not are evident. For example, the gas in simulations MH1 and MH2 is significantly less turbulent, which is reflected by the smoother appearance of the density and temperature \citep[see also][]{clark11a}. The discs in these haloes remain stable for a longer period of time, and the gas tends to be more condensed, which is evident from the smaller sizes of the discs and the protostars.

The evolution of the masses and radii of the protostars are shown in Fig.~7. Each line denotes an individual protostar, and the total mass of all protostars is denoted by the thick dotted line. Including the primary protostar, a total of 5, 10, 9, and 11 protostars form, out of which 2, 4, 5, and 5 survive until the end of the simulation. The masses of the protostars range from a few times $10^{-3}$ to almost $1\,{\rm M}_\odot$. The total mass is dominated by the primary protostar at the centre of the cloud, which is typically five times as massive as the second most massive protostar. The radii of the protostars are shown in the right panel of Fig.~7, and are in the expected range of a few tens to two hundred solar radii \citep{ho09,smith11b}. In Fig.~8, we show the total accretion rate of all protostars and the fraction that goes into the primary protostar, averaged over a $1$-yr time period. The accretion rate varies strongly between about $10^{-2}$ and $1\,{\rm M}_\odot\,{\rm yr}^{-1}$, with a mean of about $0.1\,{\rm M}_\odot\,{\rm yr}^{-1}$. More than half of the mass is accreted by the primary protostar, and the rest is distributed among the secondary protostars.

In Fig.~9, we analyse the central gas cloud as a whole as the protostellar system evolves. From top left to bottom right, the panels show the density of hydrogen nuclei, temperature, sound speed, radial velocity, rotation velocity, ratio of radial to rotation velocity, ratio of rotation to Keplerian velocity and turbulent Mach number as a function of distance to the centre of the primary protostar. The various line styles denote the output times according to the legends, and the vertical grey lines also show the radii of the primary protostars. The profiles are centred on the first protostar and have been determined according to the description in Section~3.1, but using only data from the final resimulations. The spatial range is also much smaller, since only the central regions have had enough time to evolve from their initial state. The formation of additional protostars over time is evident from the bumps in the density profiles, which show an increasing deviation from the initial profiles as the gas clouds become highly inhomogeneous. The temperature beyond the primary protostar does not increase by much, showing again that cooling within the disc is very efficient. The location of the upturn of the radial velocity, as well as the downturn of the ratio of the radial velocity to the rotation velocity, increases over time and indicates where the infall stalls and the gas becomes rotationally supported. The Keplerian velocity within this region increases to nearly unity and shows the formation and growth of a disc that encompasses the secondary protostars with their own small discs. The turbulent Mach number of the gas within this disc increases by about a factor of $2$, and is generated by the self-gravity of the gas and the non-axisymmetric nature of the system. Below about $1\,{\rm au}$, which is typically located within the protostars, the turbulent motions become subsonic.

\begin{figure}
\begin{center}
\includegraphics[width=8cm]{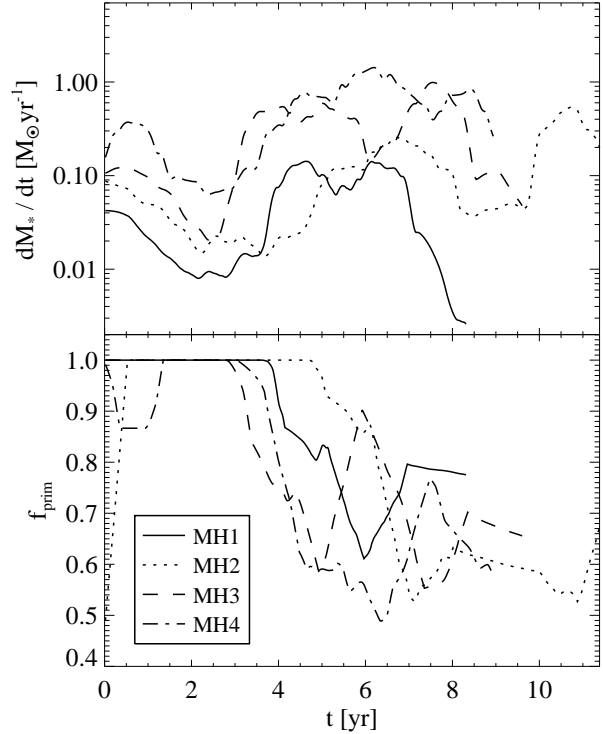}
\caption{Total accretion rate of the protostellar system (top panel), and the fraction that accretes on to the primary protostar (bottom panel) versus time after the formation of the primary protostar. The various line styles denote the minihaloes according to the legend. The accretion rates vary strongly between about $10^{-2}$ and $1\,{\rm M}_\odot\,{\rm yr}^{-1}$, with a mean of about $0.1\,{\rm M}_\odot\,{\rm yr}^{-1}$. In MH1 and MH2, HD cooling becomes important and leads to a systematically lower accretion rate. Typically more than half of the mass accretes on to the primary protostar, which grows much more rapidly than the secondary protostars.}
\end{center}
\end{figure}

\begin{figure*}
\begin{center}
\resizebox{16.5cm}{21.8cm}
{\unitlength1cm
\begin{picture}(16.5,21.8)
\put(0,11){\includegraphics[width=8cm,height=10.8cm]{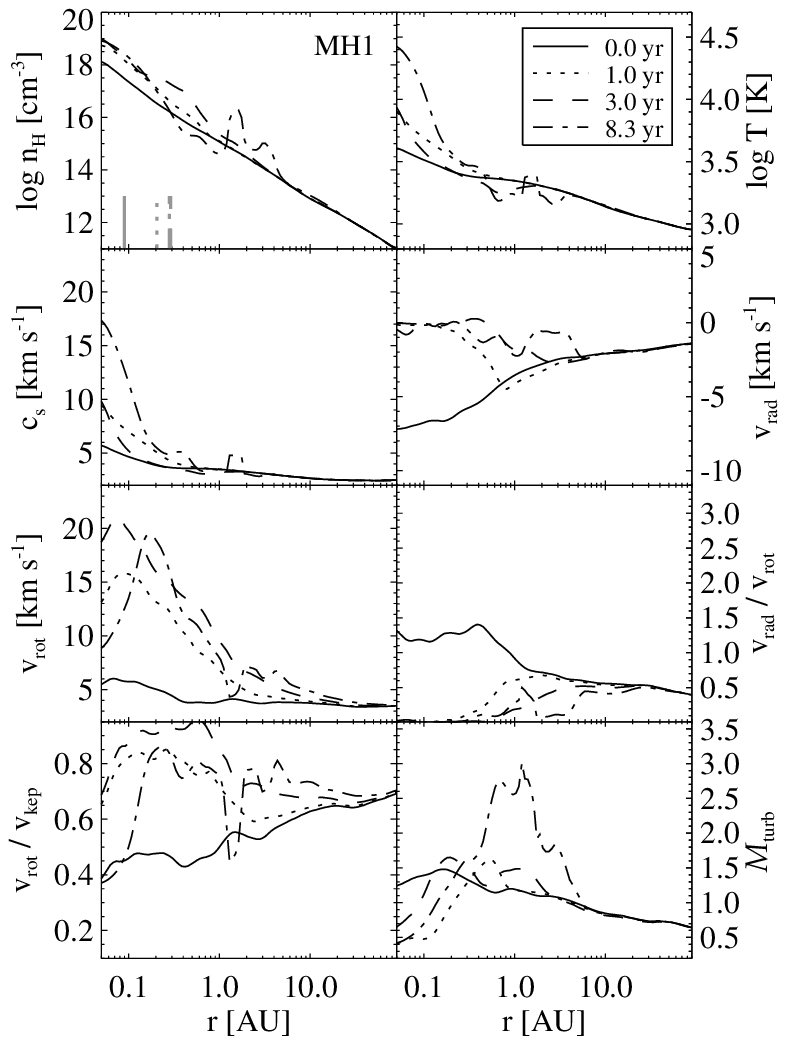}}
\put(8.5,11){\includegraphics[width=8cm,height=10.8cm]{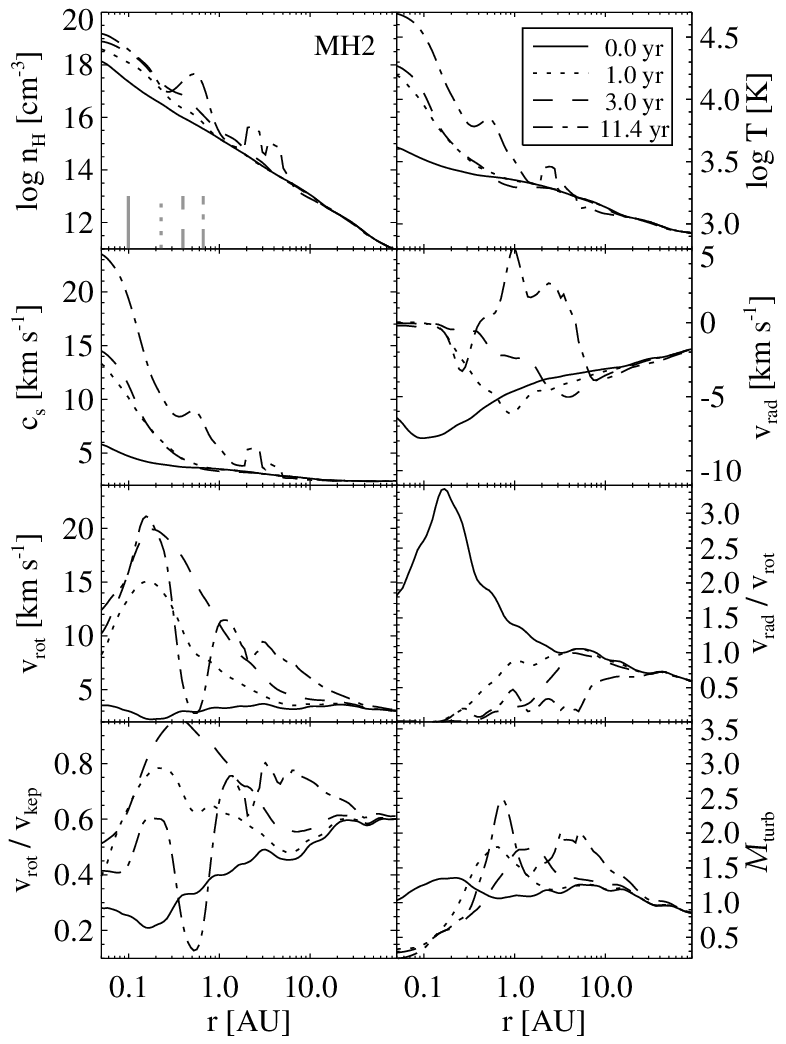}}
\put(0,0){\includegraphics[width=8cm,height=10.8cm]{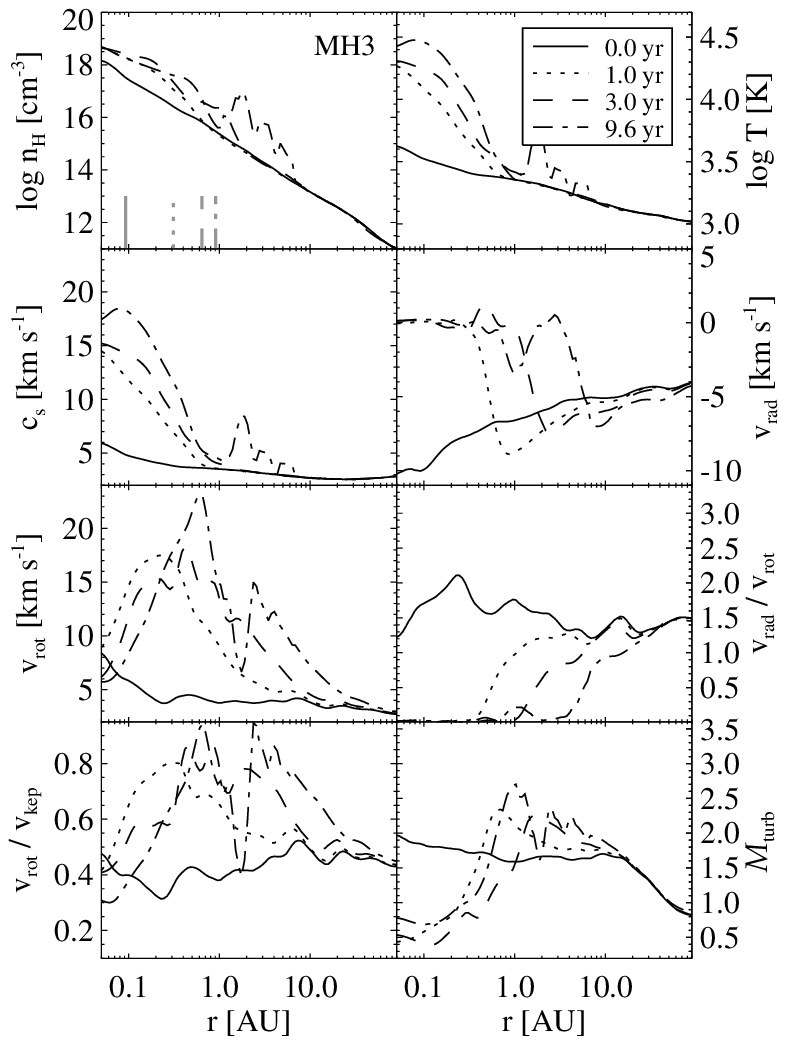}}
\put(8.5,0){\includegraphics[width=8cm,height=10.8cm]{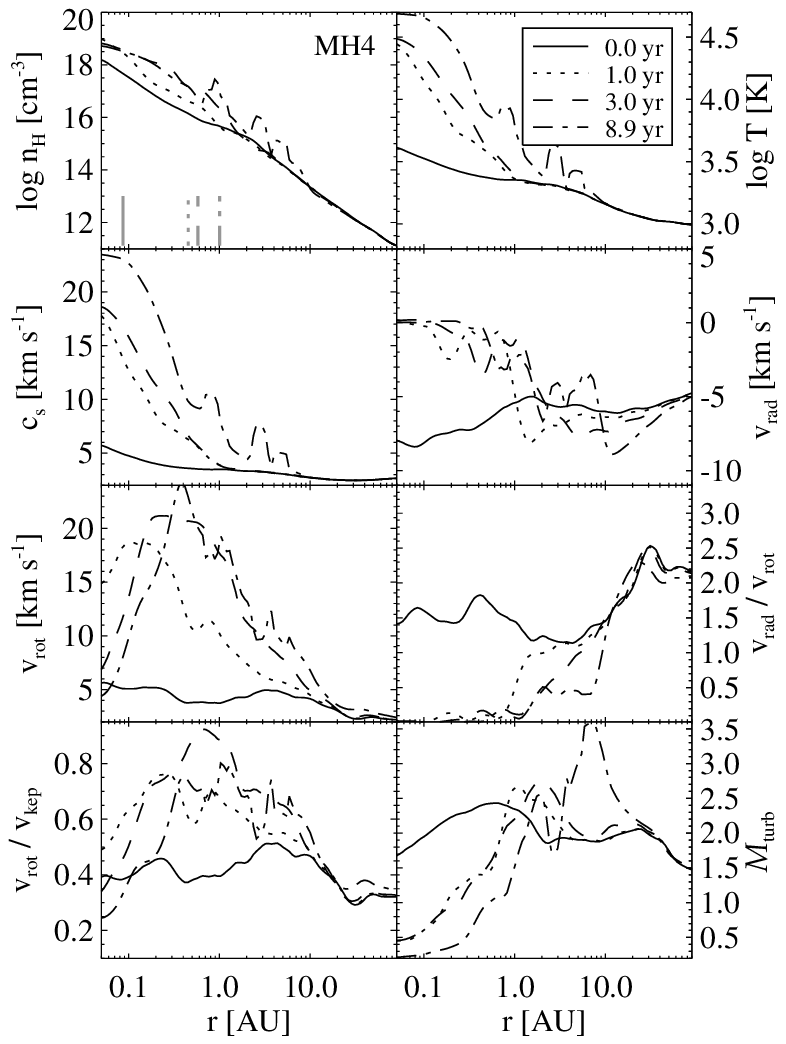}}
\end{picture}}
\caption{From top left to bottom right within each panel: density of hydrogen nuclei, temperature, sound speed, radial velocity, rotation velocity, ratio of radial to rotation velocity, ratio of rotation to Keplerian velocity, and turbulent Mach number versus distance to the centre of the primary protostar. Each panel shows an individual minihalo, and the various lines styles denote the above quantities at the output times shown in the legend. The vertical grey lines in the top left-hand panels indicate the radii of the primary protostars in each halo. This figure is discussed in detail in Section~3.3.}
\end{center}
\end{figure*}

\begin{figure}
\begin{center}
\includegraphics[width=8cm]{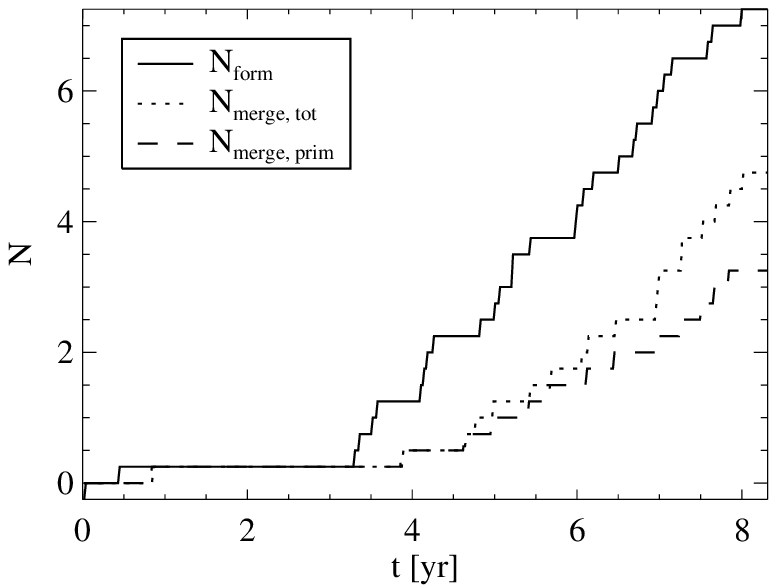}
\caption{Average number of secondary protostars formed (solid line), merged (dotted line) and merged with the primary protostar (dashed line) in each halo. At the last common output time, about half of the secondary protostars have merged with the primary protostar, and only about one third survive. Nevertheless, the multiplicity of the protostellar system increases monotonically until the end of the simulation.}
\end{center}
\end{figure}

\subsection{Merger and survival rate}

One of the key parameters that controls the stellar mass function is how efficiently the protostars merge with each other. In Fig.~10, we compare the average number of secondary protostars formed per halo with the average number of mergers. By the end of the simulation, nearly two thirds of the secondary protostars have merged away, and only one third survive. The dashed line shows that about half of the secondary protostars have merged with the primary protostar. Evidently, they shed their angular momentum very efficiently, for which either gravitational or pressure gradient torques are responsible. In the following, we investigate in turn the evolution of secondary protostars that merge with the primary protostar, and of secondary protostars that survive until the end of the simulation.

In our first attempt to determine the various torques acting on the protostars, we assumed that they may be approximated as point masses. However, this proved to be inaccurate, since the complex density and temperature profiles near the protostellar surface strongly affect the torques. This prompted us to use {\sc arepo} to generate an additional output for each snapshot that contains the gravitational acceleration and pressure gradient of all cells. We then determine the total gravitational and pressure gradient torques per unit mass, ${\bmath\tau}_{\rm grav}$ and ${\bmath\tau}_{\rm pres}$, on the protostars via summation over all cells that lie within the protostars:
\begin{equation}
{\bmath\tau}_{\rm grav}=\frac{1}{M_*}\sum_i{\bmath r}_i{\bmath\times}(m_i\,{\bmath a}_i),
\end{equation}
\begin{equation}
{\bmath\tau}_{\rm pres}=\frac{1}{M_*}\sum_i{\bmath r}_i{\bmath\times}\left(m_i\frac{\nabla P_i}{\rho_i}\right),
\end{equation}
where $i$ is the index of the cell, ${\bmath r}_i$ its distance to the centre of the primary protostar, $m_i$ its mass, $\rho_i$ its density, ${\bmath a}_i$ its gravitational acceleration, and $\nabla P_i$ its pressure gradient. We also determine the angular momentum per unit mass ${\bmath{\mathit l}}$, which is given by:
\begin{equation}
{\bmath{\mathit l}}=\frac{1}{M_*}\sum_i{\bmath r}_i{\bmath\times}\left(m_i {\bmath v}_i\right),
\end{equation}
where ${\bmath v}_i$ denotes the velocity of the cell with respect to the velocity of the primary protostar. From the angular momenta and torques, we determine the time-scales for gravitational and pressure gradient torques to operate:
\begin{equation}
 t_{\rm grav}=\frac{\left|{\bmath{\mathit l}}\right|^2}{{\bmath{\mathit l}}{\bmath\cdot}{\bmath\tau}_{\rm grav}}\quad{\rm and}\quad t_{\rm pres}=\frac{\left|{\bmath{\mathit l}}\right|^2}{{\bmath{\mathit l}}{\bmath\cdot}{\bmath\tau}_{\rm pres}}.
\end{equation}
Note that positive values indicate that the angular momentum per unit mass is increasing, while negative values indicate that it is decreasing. Finally, numerical torques are expected to be significantly smaller in {\sc arepo} than in {\sc gadget}, which employs a relatively high artificial viscosity. For the latter, \citet{sbl11} showed that numerical torques are less important than gravitational or pressure gradient torques in simulations that are similar in nature to ours, so that we are confident that we may neglect them here.

\begin{figure*}
\begin{center}
\resizebox{16cm}{14cm}
{\unitlength1cm
\begin{picture}(16,14)
\put(0,7){\includegraphics[width=8cm,height=7cm]{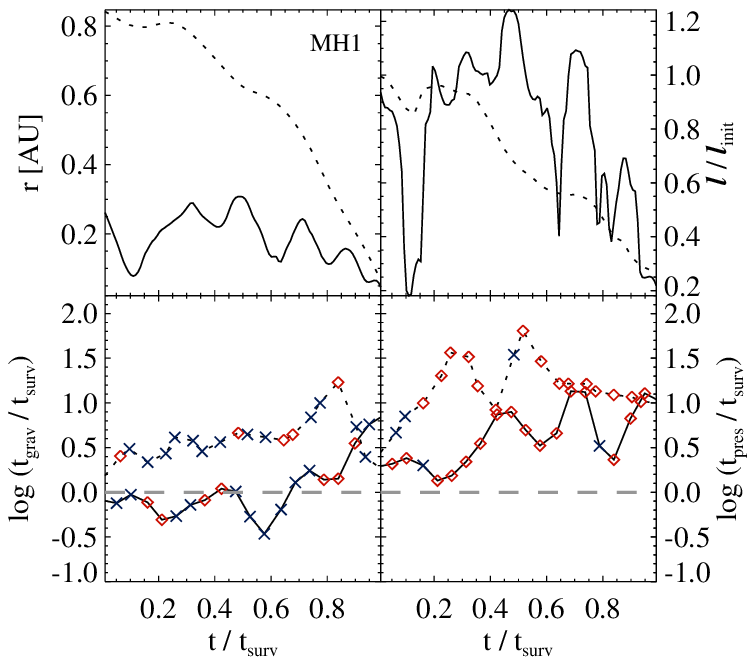}}
\put(8,7){\includegraphics[width=8cm,height=7cm]{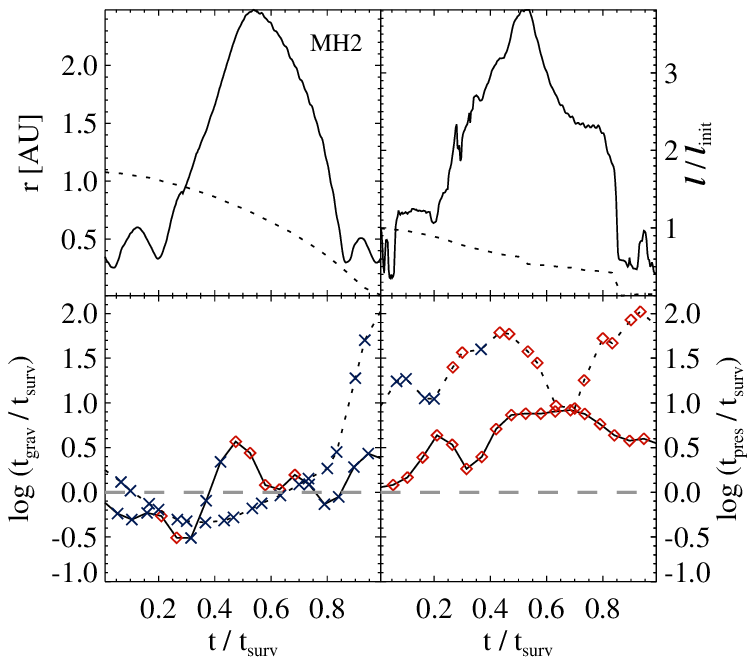}}
\put(0,0){\includegraphics[width=8cm,height=7cm]{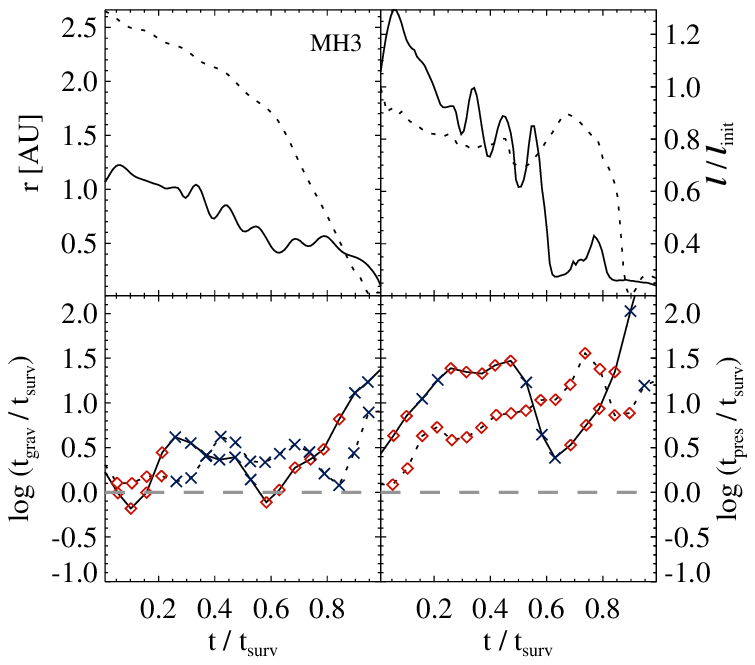}}
\put(8,0){\includegraphics[width=8cm,height=7cm]{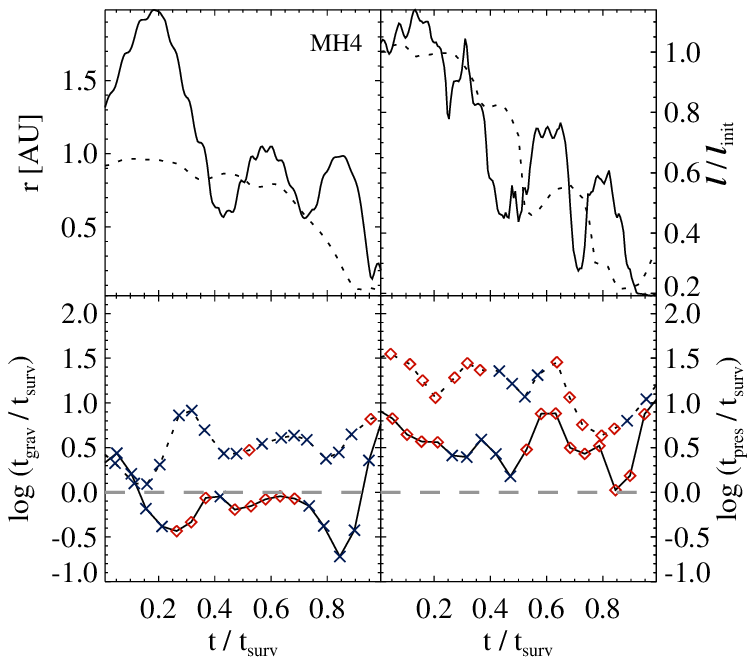}}
\end{picture}}
\caption{The evolution of the two most long-lived secondary protostars in each minihalo, which in addition merge with the primary protostar before the end of the simulation. The x-axis denotes the ratio of the time after their formation to the time that they survive. From top left to bottom right, the sub-panels show the distance to the centre of the primary protostar, the ratio of the angular momentum per unit mass to the initial value when the protostars were formed and the ratio of the time-scales for gravitational and pressure gradient torques to operate to the times that the protostars survive. The thick grey dashed lines denote a ratio of unity, and the red diamonds and blue crosses denote torques that increase and decrease the angular momentum of the protostars, respectively. The predominance of crosses in the bottom left-hand panels shows that gravitational torques tend to decrease the angular momentum of the protostars. On the other hand, torques exerted by pressure gradients typically increase the angular momentum of the protostars. The time-scales for gravitational torques to operate are generally close to the time the protostars survive, and significantly smaller than the time-scales for pressure gradient torques to operate, showing that gravitational torques are indeed responsible for the migration of the protostars to the centre of the cloud.}
\end{center}
\end{figure*}

\begin{figure*}
\begin{center}
\resizebox{16cm}{14cm}
{\unitlength1cm
\begin{picture}(16,14)
\put(0,7){\includegraphics[width=8cm,height=7cm]{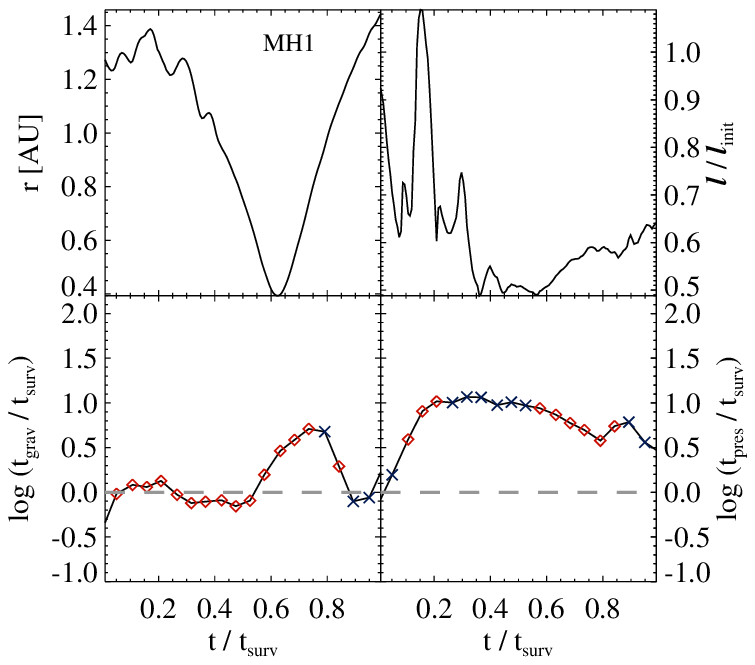}}
\put(8,7){\includegraphics[width=8cm,height=7cm]{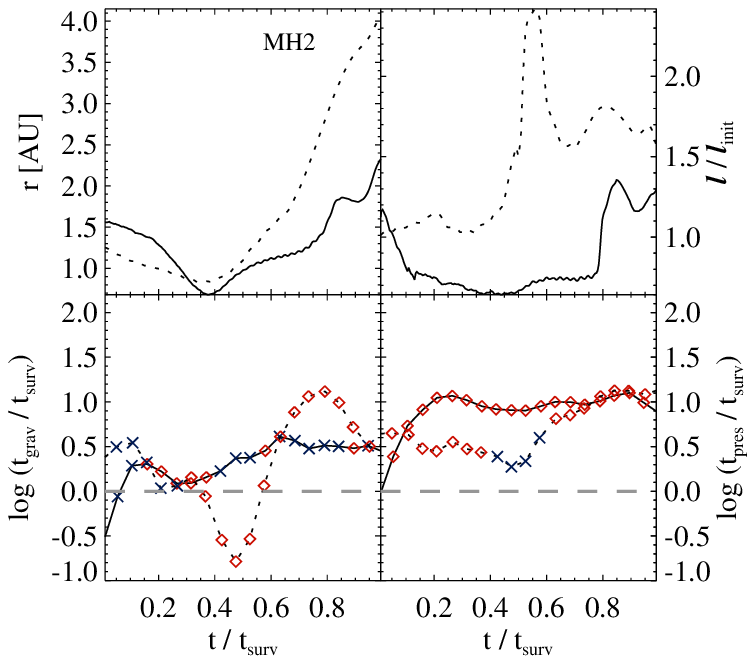}}
\put(0,0){\includegraphics[width=8cm,height=7cm]{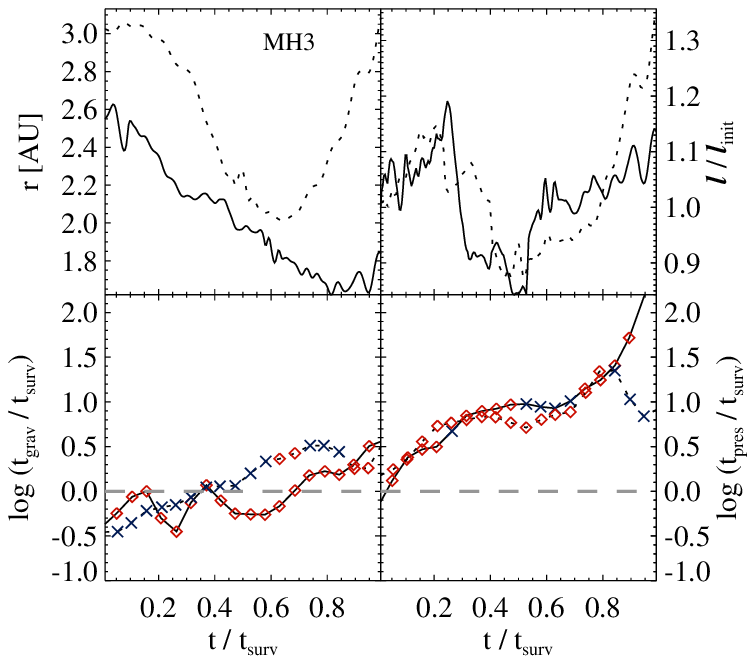}}
\put(8,0){\includegraphics[width=8cm,height=7cm]{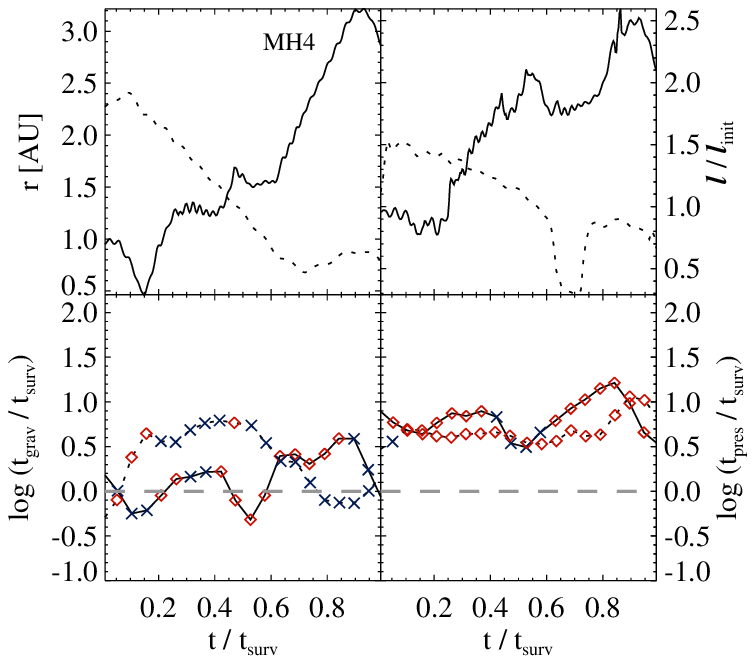}}
\end{picture}}
\caption{The evolution of the two most long-lived secondary protostars in each minihalo, which in addition survive until the end of the simulation (note that in MH1 only one secondary protostar survives). The x-axis denotes the ratio of the time after their formation to the time that they survive. From top left to bottom right, the panels show the distance to the centre of the primary protostar, the ratio of the angular momentum per unit mass to the initial value when the protostars were formed and the ratio of the time-scales for gravitational and pressure gradient torques to operate to the times that the protostars survive. The thick grey dashed lines denote a ratio of unity, and the red diamonds and blue crosses denote torques that increase and decrease the angular momentum of the protostars, respectively. In comparison with Fig.~11, diamonds tend to dominate, which indicates that gravitational torques increase the angular momentum of some of the protostars, and often results in their migration to higher orbits.}
\end{center}
\end{figure*}

In Fig.~11, we analyse the evolution of the two most long-lived secondary protostars in each halo, which in addition merge with the primary protostar. The x-axis shows the ratio of the time after their formation to the time they survive. The top left panels show the distance of the protostars to the centre of the primary protostar. The top right panels show the ratio of the angular momentum per unit mass to the initial value when the protostar is formed. Finally, the bottom panels show the ratio of the time-scales for gravitational and pressure gradient torques to operate to the times the protostars survive. Red diamonds denote a positive value, and blue crosses denote a negative value. For better visibility, we have smoothed the lines in the bottom two panels over a small period of time.

A number of important conclusions may be drawn from this figure. First, the time-scales for gravitational torques to operate are smaller than those for pressure gradients, showing that the former are more important. Secondly, the predominance of negative values in the bottom left panels shows that gravitational torques typically lead to a decrease of the angular momentum of the protostars, while the opposite is the case for the torques exerted by pressure gradients. Finally, the time-scales for gravitational torques to operate are generally close to the survival time, showing that gravitational torques are indeed responsible for the migration of the protostars to the centre of the cloud. A closer look at the evolution of individual protostars also shows that the dips and peaks in the time-scale profiles correlate well with those in the distance and angular momentum. For example, the solid line in MH2 shows a clear peak in the distance and angular momentum at about $t/t_{\rm surv}=0.5$, which is accompanied by a momentary increase of $t_{\rm grav}$ to positive values. In a second example, the distance and angular momentum of the protostar denoted by the dotted line in MH3 continuously decline, which is reflected by an extended period of negative gravitational torques. Although not entirely apparent from the bottom panels, due to the smoothing applied to the profiles, the torques acting on the protostars are also highly variable in time.

In Fig.~12, we show the evolution of the two most long-lived secondary protostars in each halo, which in addition survive until the end of the simulation (note that in MH1 only one secondary protostar survives). In this case, and in contrast to Fig.~11, gravitational torques tend to be dominated by positive values, and may lead to an increase of the distance and angular momentum of individual protostars. Two examples are the protostars denoted by the dotted line in MH2 and the solid line in MH4, which increase their separation to the centre of the primary protostar by a factor of a few. This is caused by torques from nearby protostars, indicated, for example, by the correlation of the dotted line in MH2 in Fig.~12 with the solid line in Fig.~11. The resulting migration of the protostars to higher orbits is similar to the dynamical ejections found in \citet{greif11a}, but less severe since a fraction of the angular momentum of the protostars is transferred to the surrounding gas instead of to other protostars.

In Fig.~13, we compare the average free-fall time and dynamical friction time of the secondary protostars that merge with the primary protostar to the time they survive. The free-fall time $t_{\rm ff}$ is given by:
\begin{equation}
t_{\rm ff}=\sqrt{\frac{3\pi}{32 G\rho_{\rm cloud}}},
\end{equation}
where $G$ is the gravitational constant and $\rho_{\rm cloud}$ is the mean density of the gas, which we estimate by using the mass enclosed within the distance of the protostar to the centre of the primary protostar. We also determine the dynamical friction time $t_{\rm fric}$ with respect to the gas cloud as a whole:
\begin{equation}
t_{\rm fric}=\frac{M_*\left|{\bmath v}_*\right|}{F_{\rm fric}},
\end{equation}
where $F_{\rm fric}$ is the dynamical friction force. We estimate the dynamical friction force with the approximate expression for collisionless systems:
\begin{equation}
F_{\rm fric}=\frac{4\pi G^2 M_*^2\rho_{\rm cloud}}{\left|{\bmath v}_*\right|^2},
\end{equation}
where ${\bmath v}_*$ is the velocity of the protostar with respect to the velocity of the primary protostar. We note that more sophisticated expressions exist \citep[e.g.][]{ostriker99}, but in light of the complicated morphology of the gas cloud their applicability is questionable here. The above time-scales are calculated using data from all haloes and output times, and we bin them according to the distance of the protostars to the centre of the primary protostar.

Fig.~13 shows that the reorganization of the angular momentum of the protostars and the resulting migration to the centre of the cloud typically occur in a free-fall time, which is the smallest time-scale on which a gravitationally driven process may operate. As indicated by the dashed line, dynamical friction may be partially responsible for the strong gravitational torques, in particular at small radii where the density is high. Other agents include irregularities in the gas cloud induced by turbulence or spiral arm patterns, and nearby protostars. A more sophisticated analysis of the origin of the torques would require an implementation within {\sc arepo} of the protostar finder described in Section 2.7, together with an auxiliary routine to decompose the gravitational force into its various components. However, this is beyond the scope of the present study.

A direct comparison with previous sink particle based work is problematic, since we probe a different regime in both space and time. For example, in \citet{clark11b} the first sink particle formed after $\simeq 100\,{\rm yr}$ at about $10\,{\rm au}$ from the primary protostar, while in the present study we only capture the first $10\,{\rm yr}$, and the protostars form within a few au, which is close to the accretion radius of the sink particles. The same holds for \citet{greif11a}, with an even lower resolution in other studies \citep[e.g.][]{sgb10,smith11a}. It is therefore unclear whether the processes operating on the small scales investigated here have a strong influence on the gravitational stability of the gas on larger scales. An important goal is therefore to increase the simulated timespan and allow a direct comparison with previous work.

\begin{figure}
\begin{center}
\includegraphics[width=8cm]{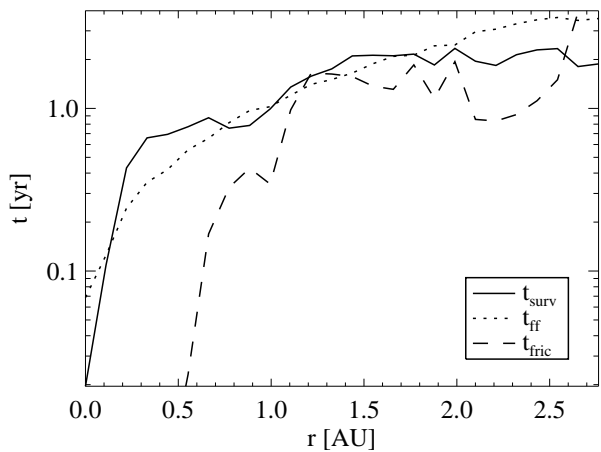}
\caption{Average survival time (solid line), free-fall time (dotted line), and dynamical friction time (dashed line) of the protostars that have merged with the primary protostar before the end of the simulation, shown as a function of distance to the centre of the primary protostar. The redistribution of angular momentum by gravitational torques is so efficient that it leads to the migration of about half of the secondary protostars to the centre of the cloud in only a free-fall time, which is the smallest time-scale on which a gravitationally driven process may operate. The dashed line shows that the torques are at least in part due to dynamical friction -- especially at small radii where the densities are high.}
\end{center}
\end{figure}

\section{Summary and conclusions}
Recent simulations of primordial star formation have suggested that the gas at the centre of minihaloes fragments into a system of protostars, as a consequence of the high accretion rate from the envelope on to the disc, and the efficient cooling of the disc by molecular hydrogen (Clark et al. 2008, 2011a,b; Stacy et al. 2010, Greif et al. 2011a). These studies employed sink particles to avoid the Courant constraint, which allowed a continuation of the simulations beyond the formation of the first protostar, at the cost of not accurately resolving the hydrodynamics near the accretion radius. We have attempted to overcome this limitation by not inserting sink particles, and instead directly resolving individual protostars and their interaction with the surrounding gas. In analogy to previous work, we find that the Keplerian disc formed around the primary protostar fragments into a number of secondary protostars, which is enabled by H$_2$ collisional dissociation cooling and collision-induced emission.

Once the first fragments have formed, aggressive gravitational torques between the protostars and the gas begin to redistribute their angular momentum. This leads to the migration of about half of the secondary protostars formed in the disc to the centre of the cloud in a free-fall time, where they merge with the primary protostar and facilitate its growth to about five times the mass of the second most massive protostar. The preferential merging with the primary protostar is akin to the `runaway growth' of a single object encountered in studies of planet formation \citep{ki96} -- as opposed to the `oligarchic growth' of multiple objects \citep{ki98} -- and to what is seen in some simulations of massive star formation in the present-day Universe \citep[e.g.][]{krumholz09}. Next to the inward migration of a large fraction of the protostars, a few also receive angular momentum from other protostars due to N-body interactions and migrate to higher orbits. Similar behavior was found in \citet{greif11a}, but with higher ejecta velocities since the angular momentum of the protostars was transferred solely to the merger product. Here, the complicated torques acting on the extended protostars results in the transfer of a fraction of the angular momentum to the surrounding gas.

The relatively high merger rate leads to the survival of only every third protostar formed in the disc. Nevertheless, the number of protostars present at any given time increases monotonically until the end of the simulation, suggesting that the protostellar system will continue to grow beyond the limited period of time simulated here. Despite the considerable computational effort involved, we have followed only a small fraction of the total accretion time, so that we cannot predict the final mass function of the first stars. In addition, we have not accurately solved for the radiation emitted by the protostars, which could significantly affect the chemical and thermal evolution of the gas (Smith et al. 2011a,b; Hosokawa et al. 2011; Stacy et al. 2012). Another complication is given by magnetic fields, which are thought to be important in minihaloes \citep[e.g.][]{mmi08,xu08,schleicher09,schleicher10,sur10,federrath11,schober12,turk12}. Next-generation simulations should therefore include additional physics, as well as continue the simulations to later times. This requires an increase of the raw computational speed of {\sc arepo}, which may be achieved by optimising time-consuming calculations such as the construction of the Voronoi tesselation, or by utilising multithreaded environments. In the near future, it may thus become possible to significantly extend the period of time that can be simulated.

\section*{Acknowledgements}
THG is indebted to Martin Asplund and Achim Weiss for granting access to their computing cluster, as well as to the Rechenzentrum Garching (RZG) and the Texas Advanced Computing Center (TACC), where most of the simulations were carried out. THG would also like to thank Kazuyuki Omukai, Masayuki Umemura, and the Institute for the Physics and Mathematics of the Universe (IPMU) in Tokyo for providing funding for an extended stay in Japan, where part of this work was completed. THG is further indebted to Tom Abel for insightful discussions during a visit at the Kavli Institute for Particle Astrophysics \& Cosmology (KIPAC) at Stanford, and to Zoltan Haiman and Matthew Turk at Columbia University, New York. THG also thanks Simon White, Mark Dijkstra, and Raul Angulo for many interesting discussions, and Thorsten Naab for proof-reading the manuscript. VB acknowledges support from NSF grant AST-1009928 and NASA ATFP grant NNX09AJ33G. PCC, SCOG, RJS, and RSK acknowledge support by contract research `Internationale Spitzenforschung II' of the Baden-W\"{u}rttemberg Stiftung (grant P- LS-SPII/18), and RSK and VS acknowledge support from DFG Research Center SFB 881, `The Milky Way System'. Finally, NY acknowledges support by the Grants-in-Aid for Young Scientists (S) 20674003 by the Japan Society for the Promotion of Science.

\end{document}